\documentclass[]{article}
\usepackage{amsmath,amssymb}
\usepackage{lmodern}
\usepackage{xcolor}
\usepackage[margin=1in]{geometry}
\usepackage{color}
\usepackage{fancyvrb}
\usepackage{graphicx}
\usepackage{hyperref}
\usepackage{booktabs}

\title{Power Studies For Two-sample Methods For Multivariate Data}
\author{Wolfgang Rolke}
\date{
  University of Puerto Rico - Mayaguez
	\newline
	\texttt{wolfgang.rolke@upr.edu} 
  \newline	
  \today	
}

\begin{document}
\maketitle

\begin{abstract}
   We present the results of a large number of simulation studies regarding the power of various nonparametric two-sample tests for multivariate data. This includes both continuous and discrete data. In general no single method can be relied upon to provide good power, any one method may be quite good for some combination of null hypothesis and alternative and may fail badly for another. Based on the results of these studies we propose a fairly small number of methods chosen such that for any of the case studies included here at least one of the methods has good power. 
	The studies were carried out using the R package  \emph{MD2sample}, available from CRAN. 
\end{abstract}

\section{Introduction}\label{introduction}

The nonparametric two-sample problem
has a history going back a century, with many contributions by some of
the most eminent statisticians. In a nonparametric two-sample problem we have a data set \((x_1,..,x_n)\) from some distribution \(F\) and a data set \((y_1,..,y_m)\)
from some distribution \(G\), and we want to test \(H_0:F=G\), that
is we want to test whether the two data sets were generated by the same
(unspecified) distribution.

The literature on this problem is vast and steadily growing.
Detailed discussions can be found in \cite{agostini1986},
\cite{thas2010}, \cite{raynor2009}. For an introduction to
Statistics and hypothesis testing in general see \cite{casella2002} or \cite{bickel2015}.

The power studies in this article were carried out using \textbf{R}
programs in the package \emph{MD2sample}, available from
the CRAN website. Some tests are already implemented in other packages and some of the tests have not had 
an implementation before. However, there are no packages that bring together as many tests as \emph{MD2sample}. Also

\begin{itemize}
\item
  many methods are implemented for both continuous and discrete
  data. Discrete data includes the case of histogram (aka discretized or binned) data.\\
\item
  many methods are implemented using both \emph{Rcpp} \cite{rcpp2024} and
  parallel programming.\\
\item
  some methods make use of large-sample formulas, therefore allowing for very large data sets.\\
\item
  there are routines that allow the user to combine several tests and find a corrected p value.\\
\item
  the routines can also use any other user-defined tests.\\
\item
  the packages include routines to easily carry out power studies.
\end{itemize}

\section{Types of case studies}

The cases included in this study fall into the following categories: the data is either 
two or five dimensional and continuous, or two-dimensional and discrete. Furthermore the distributions 
are chosen such that the marginal distributions are either equal or not. In the case
where the marginal distributions are the same running univariate two-sample tests would of course
not detect any difference between them.

The cases for discrete data are the same as those for continuous data in two dimensions, with the data binned.

Multivariate distributions with marginals that have a uniform $[0,1]$ distribution are known as copulas. The restriction to uniform marginals is superficial, as one can always use the probability integral transform
to turn any problem with any marginal distributions into one with uniform marginals, and vice versa.

Another worthwhile comment is as follows: if we want to restrict ourselves to the cases where two data sets have marginals that are uniform $[0,1]$, then any difference between them has to be of a very specific type. They have to differ by a shape that goes along either of the two diagonal axes. Moreover the density conditional on any interval within $[0,1]$ has to be the same as well. As an example consider figure 1, where the points shown in red are such that they do not change the marginal uniform distribution. As this is a rather specific pattern one would expect that in real data differences between the two data sets will more often than not appear already in one of the marginals and be detectable by a univariate two-sample test. In higher-dimensional data sets the situation will of course become more difficult. Still, requiring the one-dimensional marginals to be the same in the two data sets strongly limits how they can differ.

\begin{figure}[!htbp]
\centering
\includegraphics[width=4in]{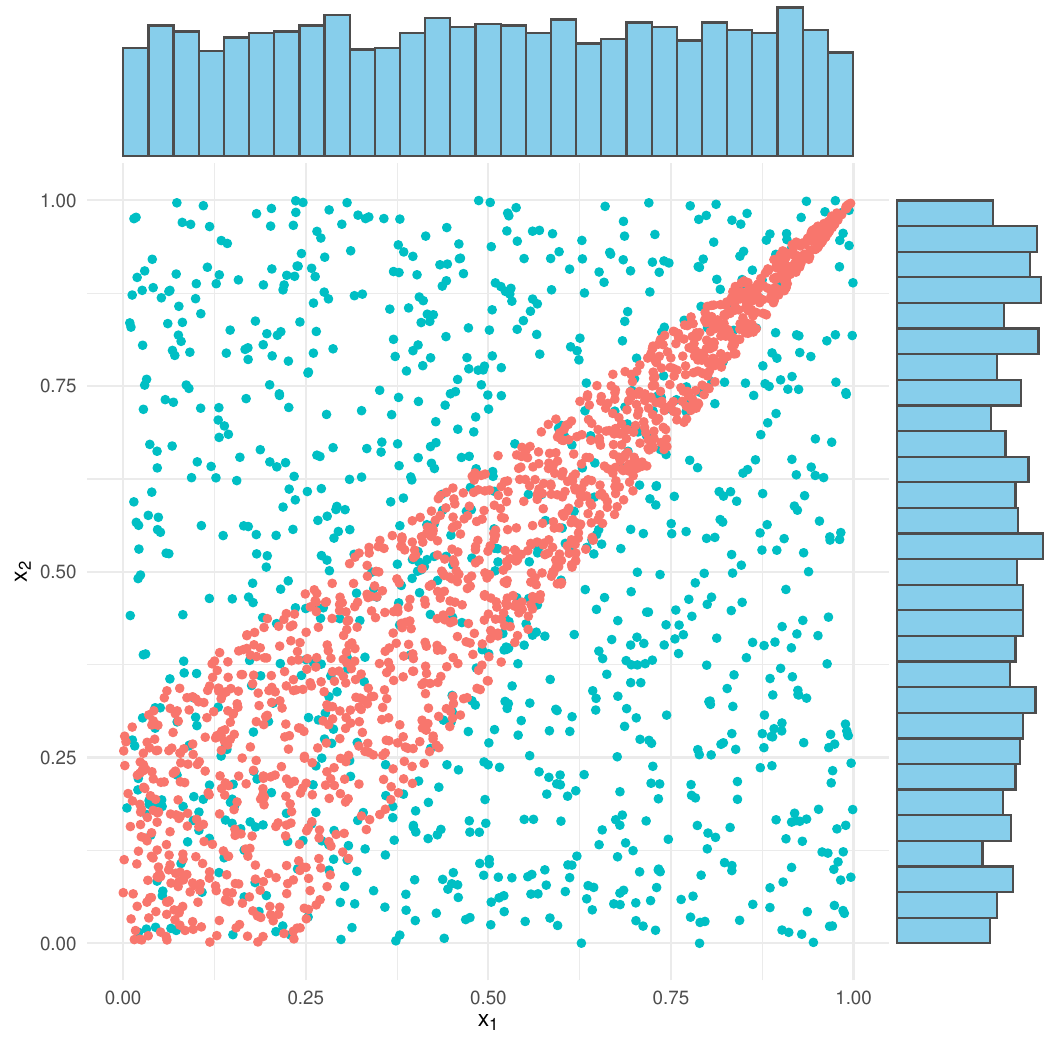}
\caption{Illustration of possible difference for Copulas. The observations in red, belonging to just one of the two data sets, need to have uniform marginals so that the two data sets overall also have uniform marginals.}
\end{figure}

For detailed explanations of the included case studies see the appendix.

\section{Results}

In this section we present the main results of this study. For the exact power estimates of each simulation study see the tables in the appendix. 

\subsection{Continuous Data}

We begin by showing the mean power of each test, sorted from highest to lowest:

\begin{table}[ht]
\centering
\begin{tabular}{c|c|c|c|c|c|c|c|c}\toprule
AZ &ES &EP &BF &AD &NN5 &CvM &KS &Ball\\
82.6 &80.8 &76.1 &71.5 &64.5 &63.2 &58.1 &53.8 &53\\\midrule
FR &CF1 &CF3 &K &CF4 &NN1 &CF2 &NN0 &BG\\
52.9 &52.9 &52.9 &50.2 &49.7 &43.9 &43.4 &36.1 &31.8\\\bottomrule
\end{tabular}
\caption{Mean power of tests for continuous data}
\end{table}

So we find that the Aslan-Zech energy test is best overall, with a mean power of $82.6\%$. A chi-square test with 5x5 bins is next, followed by the Baringhaus-Franz test. Note that in many of the case studies the marginals have a uniform distribution, so there is no real difference between equal space and equal probability bins.

The next table shows the percentage of studies where a method was close (that is within $90\%$) of the best method for each case study:

\begin{table}[ht]
\centering
\begin{tabular}{c|c|c|c|c|c|c|c|c}\toprule
AZ &ES &EP &BF &AD &NN5 &CvM &Ball &BG\\
74 &44 &38 &32 &30 &30 &26 &22 &16\\\midrule
K &FR &CF1 &CF3 &KS &CF2 &CF4 &NN1 &NN0\\
14 &14 &14 &14 &12 &6 &6 &2 &2\\\bottomrule
\end{tabular}
\caption{Percentage of times a test is close to best  for continuous data}
\end{table}

So in almost 3 out of 4 case studies the Aslan-Zech test was as good or almost so as the best method. This is much better than even the second best, a chi square test with either equal size or with equal probability bins, which only was close to best in 4 out of any 10 case studies.

Finally we find a small selection of tests in such a way that for each case study the selection includes at least one test whose power is at least $90\%$ of the best method for this case. For continuous data we find that this selection includes the Aslan-Zech (AZ) test, a chi square test with a small number of bins (ES, here 5x5), the Anderson-Darling (AD) test, the five nearest neighbor (NN5) test and finally the Biswas-Ghosh (BG) test. It is interesting to note that the Biswas-Ghosh test is in this selection even though it has the lowest mean power of all the included tests. This is likely due to the fact that the Biswas-Ghosh test is sensitive to a departure from the null hypothesis that is rare and that none of the other tests is able to detect.

It should be noted that a test not included in this short list need not be bad, but that it simply doesn't add substancially to the combined power. This is likely due to the fact that is is highly correlated with a test already included. Also, this list is not unique. 

\subsection{Discrete Data}

The results for discrete data are

\begin{table}[ht]
\centering
\begin{tabular}{c|c|c|c|c|c|c|c}\toprule
Chisquare &NN &AD &K &KS &CvM &AZ &BF\\
79.4 &53.2 &49.2 &47.4 &43.5 &42.9 &26.5 &20.2\\\bottomrule
\end{tabular}
\caption{Mean power of tests for discrete data}
\end{table}

\begin{table}[ht]
\centering
\begin{tabular}{c|c|c|c|c|c|c|c}\toprule
Chisquare &K &KS &AD &NN &AZ &CvM &BF\\
85 &21 &15 &15 &15 &15 &9 &6\\\bottomrule
\end{tabular}
\caption{Percentage of times a test is close to best  for discrete data}
\end{table}
 
Herr the small selection again includes the Aslan-Zech (AZ) test, a chi square test (Chisquare) and the Cramer-vonMises (CvM) test. In addition we need the Kolmogorov-Smirnov (KS) test.

The results for discrete data should be taken with some care. In all the case studies (for two-dimensional data) the continuous data was first binned into 5-by-5 equal size bins. This of course will greatly effect the performance of the methods, and some like the Aslan-Zech test might have much better power for other binning schemes.

\section{Methods included in this study}

\subsection{Continuous data}

\subsubsection{Methods whose p values are found via
simulation}\label{methods-whose-p-values-are-found-via-simulation}

\paragraph{Methods based on empirical distribution
functions}\label{methods-based-on-empirical-distribution-functions}

Denote by \(\mathbf{z}\) the combined data set \(x_1,..,x_n,y_1,..y_m\).
Let \(\hat{F}\) and \(\hat{G}\) be the empirical distribution functions
of the two data sets, and let \(\hat{H}\) be the empirical distribution
function of \(\mathbf{z}\)

\textbf{Kolmogorov-Smirnov test}

This classic test uses the test statistic

\[\max\{\vert \hat{F}(z_i)-\hat{G}(z_i)\vert;z_1,..,z_{n+m}\}\]

It was first proposed in (Kolmogorov 1933) and (Smirnov 1939).

\textbf{Kiuper's test}

A variant of Kolmorogov-Smirnov:

\[\max\{\hat{F}(z_i)-\hat{G}(z_i);z_1,..,z_{n+m}\}-\min\{ \hat{F}(z_i)-\hat{G}(z_i);z_1,..,z_{n+m}\}\]
This test was first discussed in (Kuiper 1960).

\textbf{Cramer-vonMises test}

\[\sum_{i=1}^{n+m} \left(\hat{F}(z_i)-\hat{G}(z_i)\right)^2\] the
extension to the two-sample problem of the Cramer-vonMises criterion is
discussed in (T. W. Anderson 1962).

\textbf{Anderson-Darling test}

\[\sum_{i=1}^{n+m} \frac{\left(\hat{F}(z_i)-\hat{G}(z_i)\right)^2}{\hat{H}(z_i)(1-\hat{H}(z_i))}\]
It was first proposed in (Theodore W. Anderson, Darling, et al. 1952).

\paragraph{Methods based on nearest
neighbors}\label{methods-based-on-nearest-neighbors}

The test statistics are the average number of nearest neighbors of the
\(\mathbf{x}\) data set that are also from \(\mathbf{x}\) plus the
average number of nearest neighbors of the \(\mathbf{y}\) data set that
are also from \(\mathbf{y}\). \emph{NN1} uses one nearest neighbor and
\(NN5\) uses 5.

\paragraph{Methods based on distances between
observations}\label{methods-based-on-distances-between-observations}

We denote by \(||.||\) Euclidean distance

\textbf{Aslan-Zech test}

This test discussed in (Aslan and and 2005) uses the test statistic

\[
\begin{aligned}
&\frac{1}{nm}\sum_{i=1}^n \sum_{j=1}^m \log(||x_i-y_j||) -\\
&\frac{1}{n^2}\sum_{i=1}^n \sum_{i<j} \log(||x_i-x_j||) - \\
&\frac{1}{m^2}\sum_{i=1}^m \sum_{i<j} \log(||y_i-y_j||) 
\end{aligned}
\] \textbf{Baringhaus-Franz test}

Similar to the Aslan-Zech test, it uses the test statistic

\[
\begin{aligned}
&\frac{nm}{n+m}\left[\frac{1}{nm}\sum_{i=1}^n \sum_{j=1}^m \sqrt{||x_i-y_j||} + \right.\\
&\frac{1}{n^2}\sum_{i=1}^n \sum_{i<j} \sqrt{||x_i-x_j||} -\\
&\left. \frac{1}{m^2}\sum_{i=1}^m \sum_{i<j} \sqrt{||y_i-y_j||} \right]\\
\end{aligned}
\] and was first proposed in (Baringhaus and Franz 2004).

\textbf{Biswas-Ghosh test}

Another variation of test based on Euclidean distance was discussed in
(Biswas and Ghosh 2014).

\[
\begin{aligned}
&B_{xy} = \frac{1}{nm}\sum_{i=1}^n \sum_{j=1}^m \sqrt{||x_i-y_j||} \\
&B_{xx}= \frac{2}{n(n-1)}\sum_{i=1}^n \sum_{i<j} \sqrt{||x_i-x_j||} \\
&B_{yy}=\frac{2}{m(m-1)}\sum_{i=1}^m \sum_{i<j} \sqrt{||y_i-y_j||}\\
&\left(B_{xx}-B_{xy}\right)^2+\left(B_{yy}-B_{xy}\right)^2
\end{aligned}
\]

\subsubsection{Methods whose p values are found using a large sample
theory}\label{methods-whose-p-values-are-found-using-a-large-sample-theory}

\textbf{Chi-square tests}

The classic tests require that the data first be binned. Because of the curse of dimensionality binning is not
likely to work in anything beyond two dimensions, and so this test is is only implemented for 2-D data. In that case the region with the data is tiled into n bins in the x direction and m bins in the y direction. The bins are either equal size or equal probability bins. In the case studies discussed here 5-by-5 bins were used.  Then these bins are combined until each contains at least 5 observations from the combined data set. 

Using only 25=5-by-5 bins might seem a bit low, but it has been shown in the past that the chi-square two-sample test has its highest power for a relatively small number of bins, regardless of the sample size, see for example \cite{rolke2021}.

\textbf{Friedman-Rafski test}

This test is a multi-dimensional extension of the classic Wald-Wolfowitz
test bases on minimum spanning trees. It was discussed in (Friedman and
Rafsky 1979).

\textbf{Simple-Nearest-Neighboor test}

Similar to the nearest neigboor tests described earlier, it uses only
the number of nearest neighbors of the first data set that are also from
the first data set. This number has a binomial distribution, and this
can be used to find p values.

\textbf{Chen-Friedman tests}

These tests, discussed in (Chen and Friedman, n.d.), are implemented in
the \emph{gTests} (Chen and Zhang 2017) package.

\textbf{Ball Divergence test}

A test described in (Pan et al. 2018) and implemented in the R package
(Zhu et al. 2021).

\subsection{Discrete data}

Implemented for discrete data are versions of the Kolmogorov-Smirnov,
Kuiper, Cramer-vonMises, Anderson-Darling, nearest neighboor,
Aslan-Zech, Baringhaus-Franz tests as well as a chisquare test.

\section{List of case studies} 

In this section we list and illustrate the case studies included in this work. Some of them are variations of Dalitz plots, a type of two dimensional data often encountered in high energy physics experiments. See for example 
\cite{asner2004}.

Case studies 1-17 with names ending in D2 are case studies where the marginal distributions are the same for the two data sets, and studies 18-34 with names ending in M are cases were at least some of the marginal distributions are not the same.

\subsection{Two dimensional data}

\subsubsection{Case Study 1: Multivariate Normal Distribution}
\begin{itemize}
\item
x: data comes from a bivariate normal distribution with mean vector $(0,0)$ and variance-covariance matrix the identity matrix.
\item
y: data comes from a bivariate normal distribution like x but with $var(y_1)=var(y_2)=1$ and $cov(y_1,y_2)=s$.
\end{itemize}
 
\begin{figure}[!htbp]
\centering
\includegraphics[width=4in]{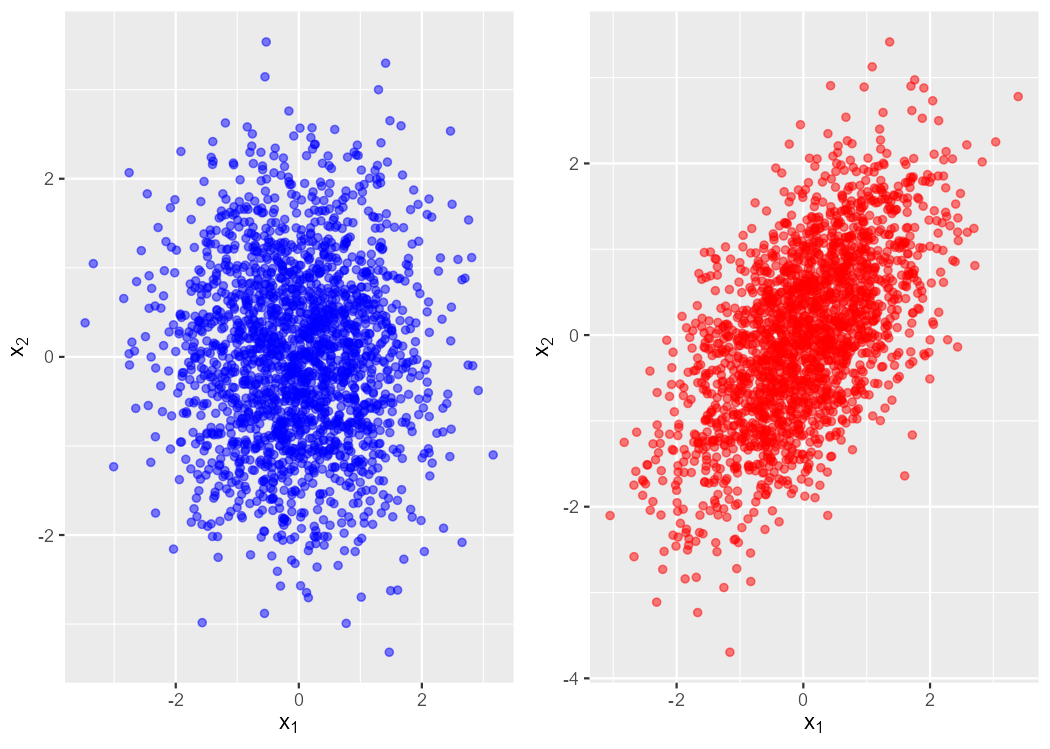}
\caption{Case Study \#1: NormalD2 (equal marginals).}
\end{figure}
\newpage
\subsubsection{Case Study 2: Multivariate t Distribution}
\begin{itemize}
\item
x: data comes from a bivariate t distribution with 5 degrees of freedom, mean vector $(0,0)$ and variance-covariance matrix the identity matrix.
\item
y: data comes from a bivariate t distribution with 5 degrees of freedom like x but with $var(y_1)=var(y_2)=1$ and $cov(y_1,y_2)=s$.
\end{itemize}
 
\begin{figure}[!htbp]
\centering
\includegraphics[width=4in]{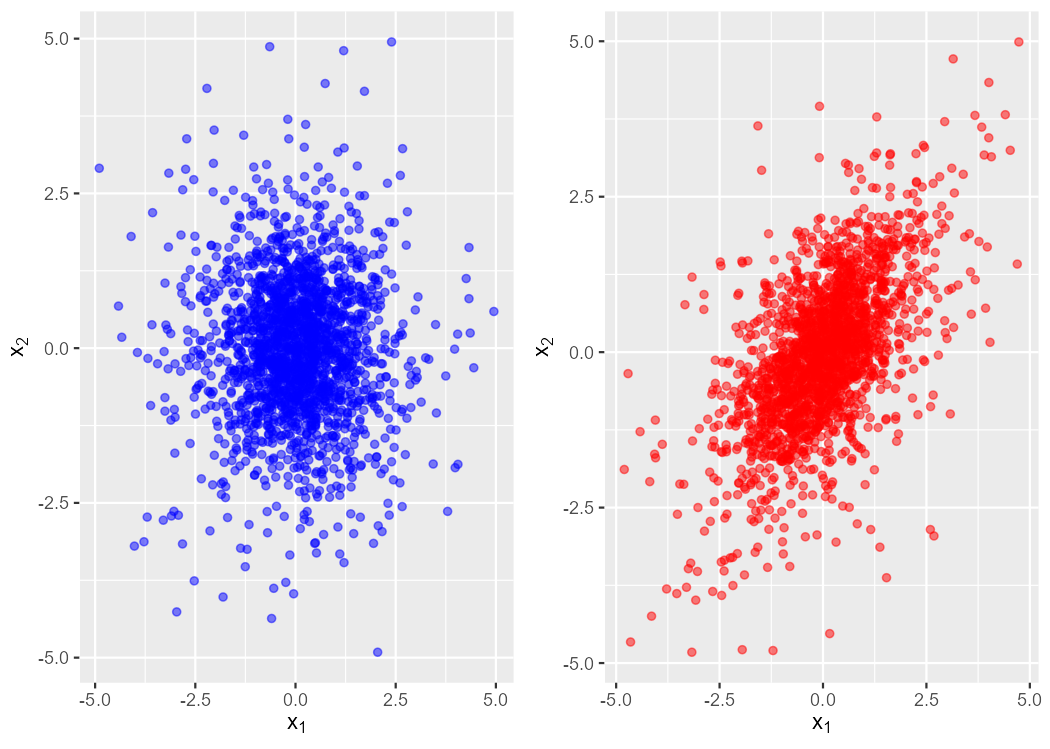}
\caption{Case Study \#2: tD2 (equal marginals).}
\end{figure}
\newpage
\subsubsection{Case Study 3: Mixture of Uniform Distributions}
\begin{itemize}
\item
x: two independent uniform random variables.
\item
y: two independent uniform random variables plus additional observations on the diagonal.
\end{itemize}
 
\begin{figure}[!htbp]
\centering
\includegraphics[width=4in]{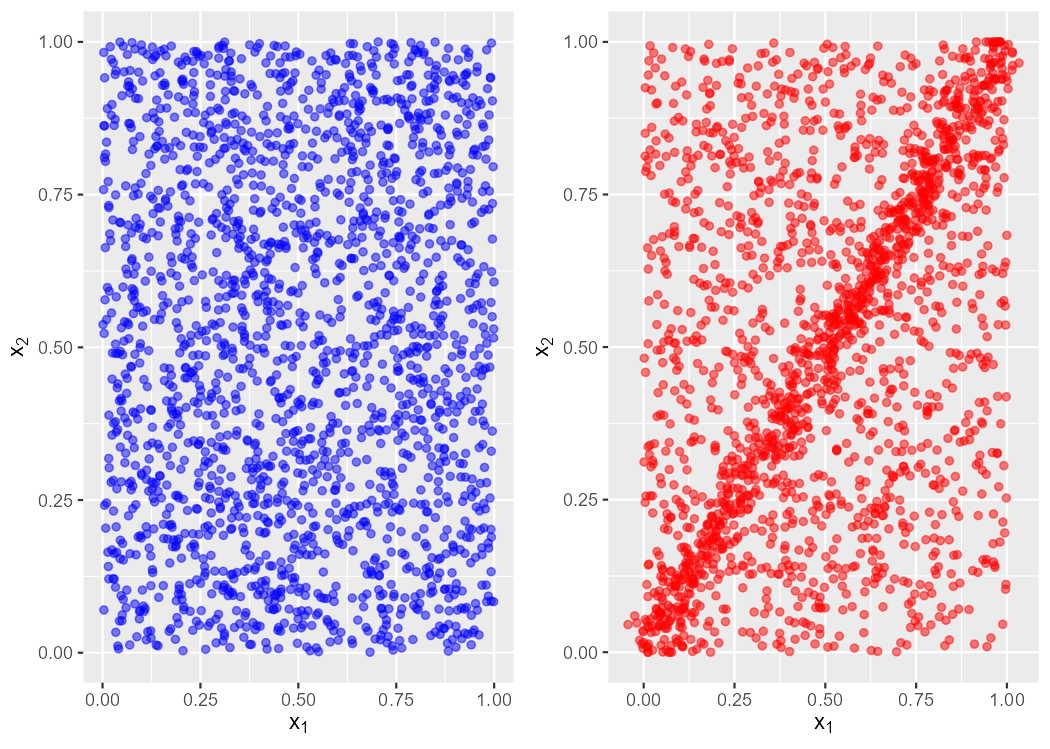}
\caption{Case Study \#3: UniformMixtureD2 (equal marginals).}
\end{figure}
\newpage
\subsubsection{Case Study 4: Frank copula}
\begin{itemize}
\item
x: two independent uniform random variables.
\item
y: observations from a Frank copula.
\end{itemize}
 
\begin{figure}[!htbp]
\centering
\includegraphics[width=4in]{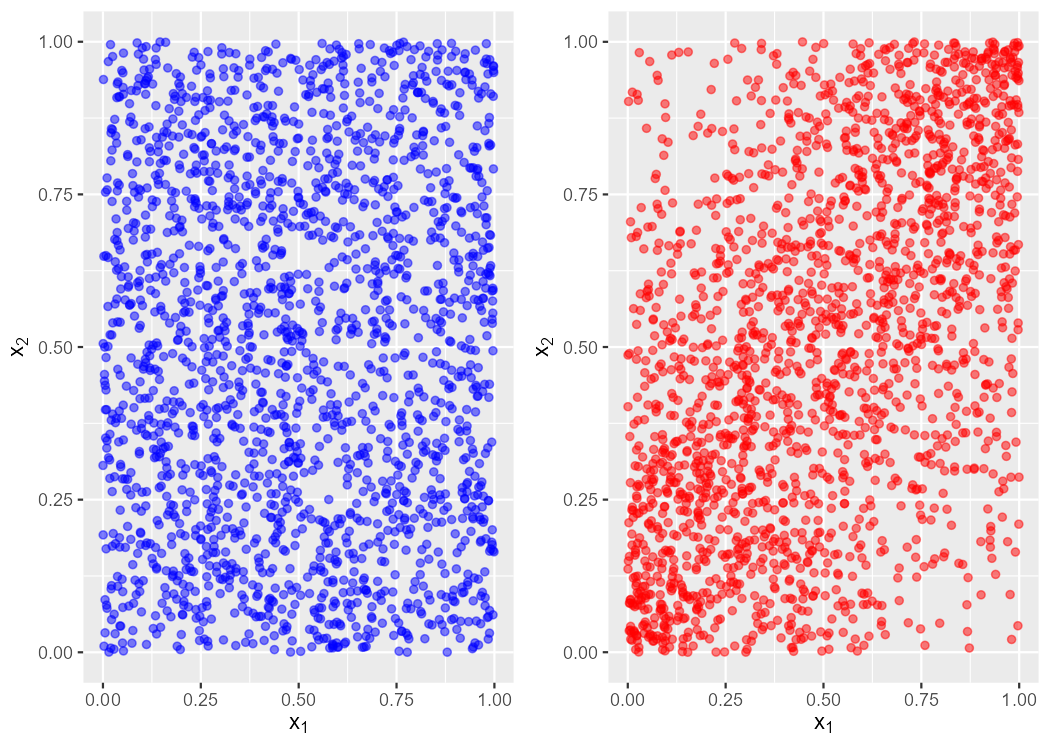}
\caption{Case Study \#4: FrankD2 (equal marginals).}
\end{figure}
\newpage
\subsubsection{Case Study 5: Clayton copula}
\begin{itemize}
\item
x: two independent uniform random variables.
\item
y: observations from a Clayton copula.
\end{itemize}
 
\begin{figure}[!htbp]
\centering
\includegraphics[width=4in]{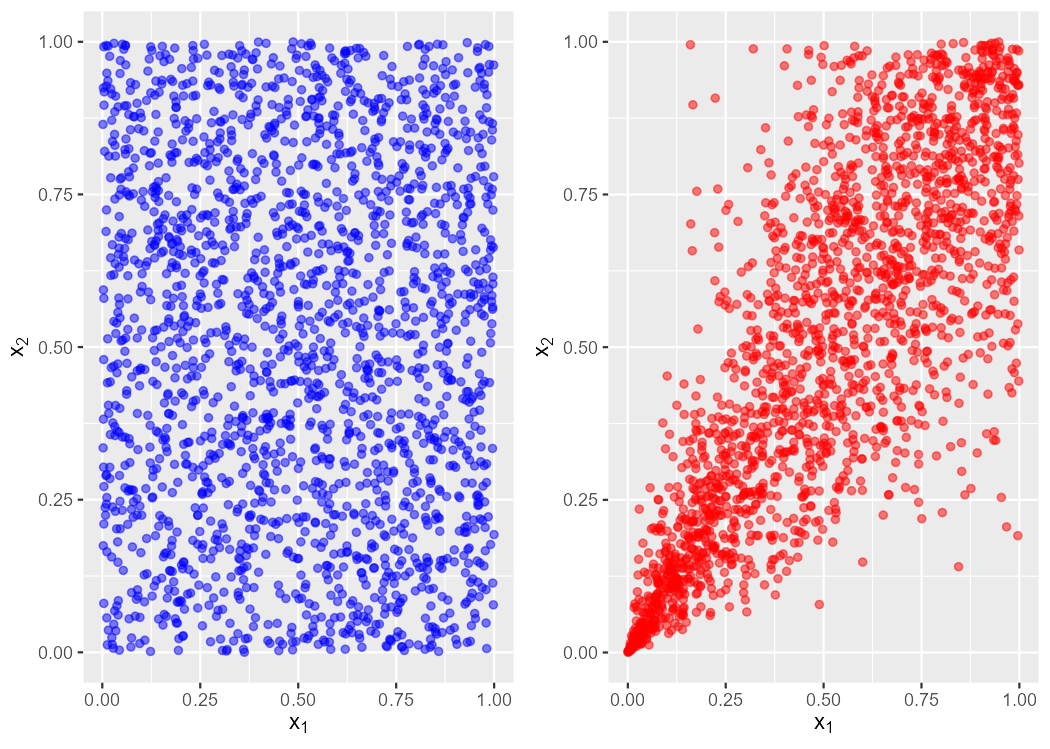}
\caption{Case Study \#5: ClaytonD2 (equal marginals).}
\end{figure}
\newpage
\subsubsection{Case Study 6: Gumbel copula}
\begin{itemize}
\item
x: two independent uniform random variables.
\item
y: observations from a Gumbel copula.
\end{itemize}
 
\begin{figure}[!htbp]
\centering
\includegraphics[width=4in]{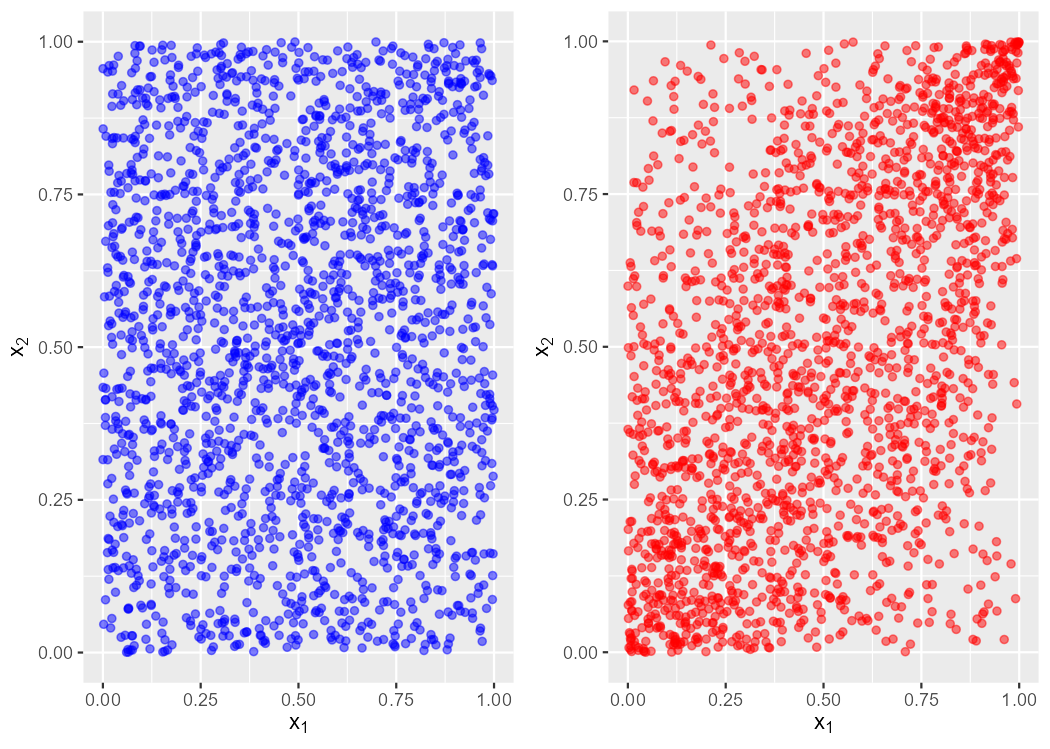}
\caption{Case Study \#6: GumbelD2 (equal marginals).}
\end{figure}
\newpage
\subsubsection{Case Study 7: Galambos copula}
\begin{itemize}
\item
x: two independent uniform random variables.
\item
y: observations from a Galambos copula.
\end{itemize}
 
\begin{figure}[!htbp]
\centering
\includegraphics[width=4in]{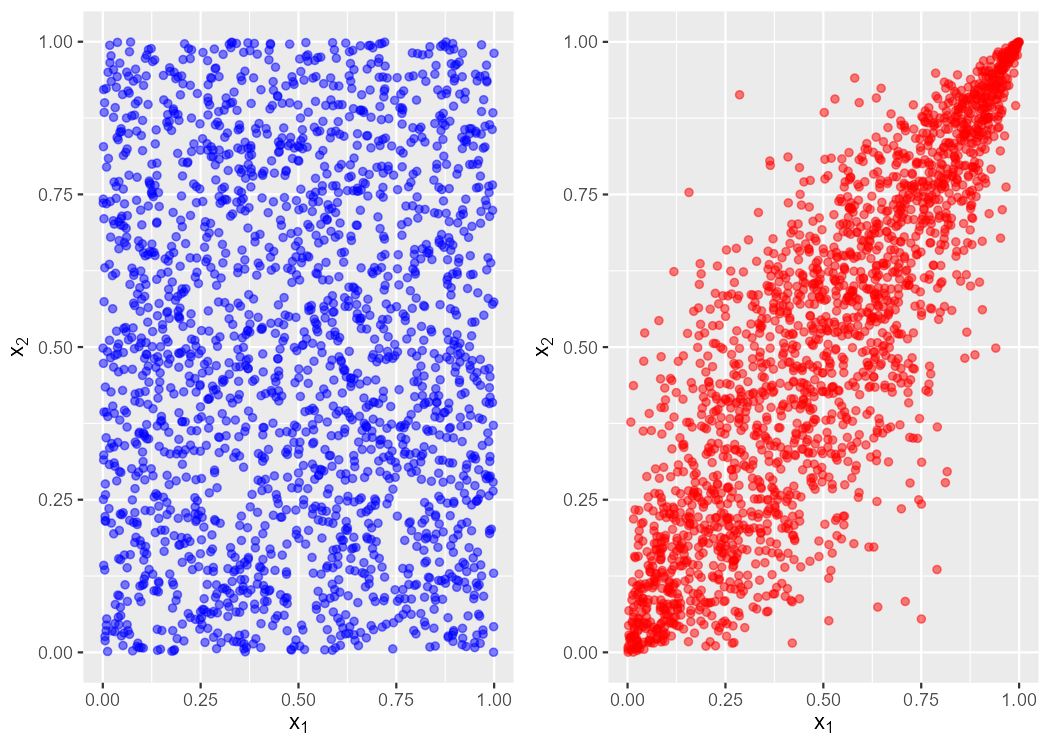}
\caption{Case Study \#7: GalambosD2 (equal marginals).}
\end{figure}
\newpage
\subsubsection{Case Study 8: Husler-Reiss copula}
\begin{itemize}
\item
x: two independent uniform random variables.
\item
y: observations from a Husler-Reiss copula.
\end{itemize}
 
\begin{figure}[!htbp]
\centering
\includegraphics[width=4in]{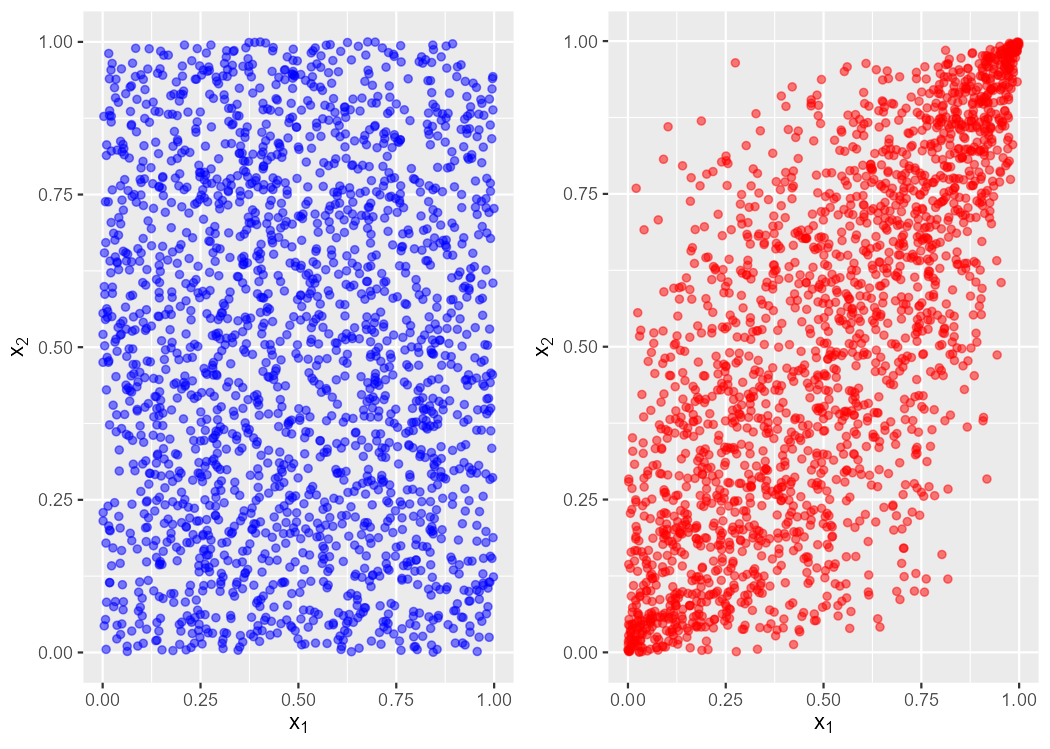}
\caption{Case Study \#8: HuslerReissD2 (equal marginals).}
\end{figure}
\newpage
\subsubsection{Case Study 9: Mixture of Clayton and Gumbel copulas}
\begin{itemize}
\item
x: observations from a 50-50 mixture of Clayton and Gumbel copulas.
\item
y: observations from a $(\alpha, 1-\alpha)$ mixture of Clayton and Gumbel copulas.
\end{itemize}
 
\begin{figure}[!htbp]
\centering
\includegraphics[width=4in]{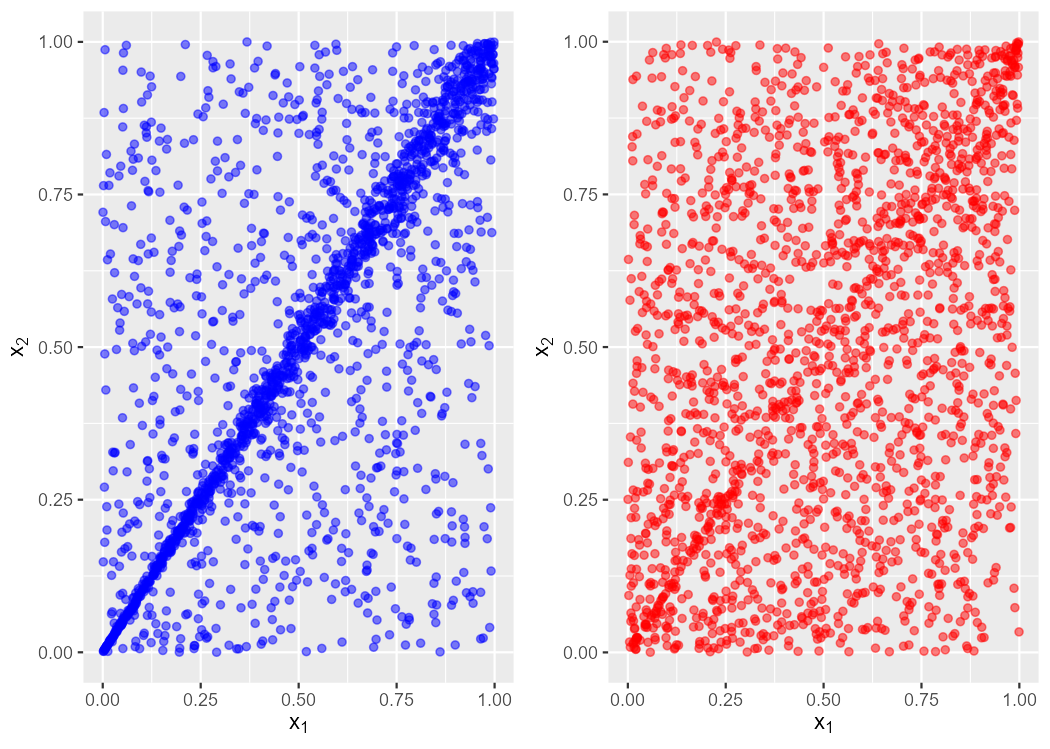}
\caption{Case Study \#9: ClaytonGumbelD2 (equal marginals).}
\end{figure}
\newpage
\subsubsection{Case Study 10: Mixture of Uniform and Frank copulas}
\begin{itemize}
\item
x: observations from a 50-50 mixture of Uniform and Frank copulas.
\item
y: observations from a $(\alpha, 1-\alpha)$ mixture of Uniform and Frank copulas.
\end{itemize}
 
\begin{figure}[!htbp]
\centering
\includegraphics[width=4in]{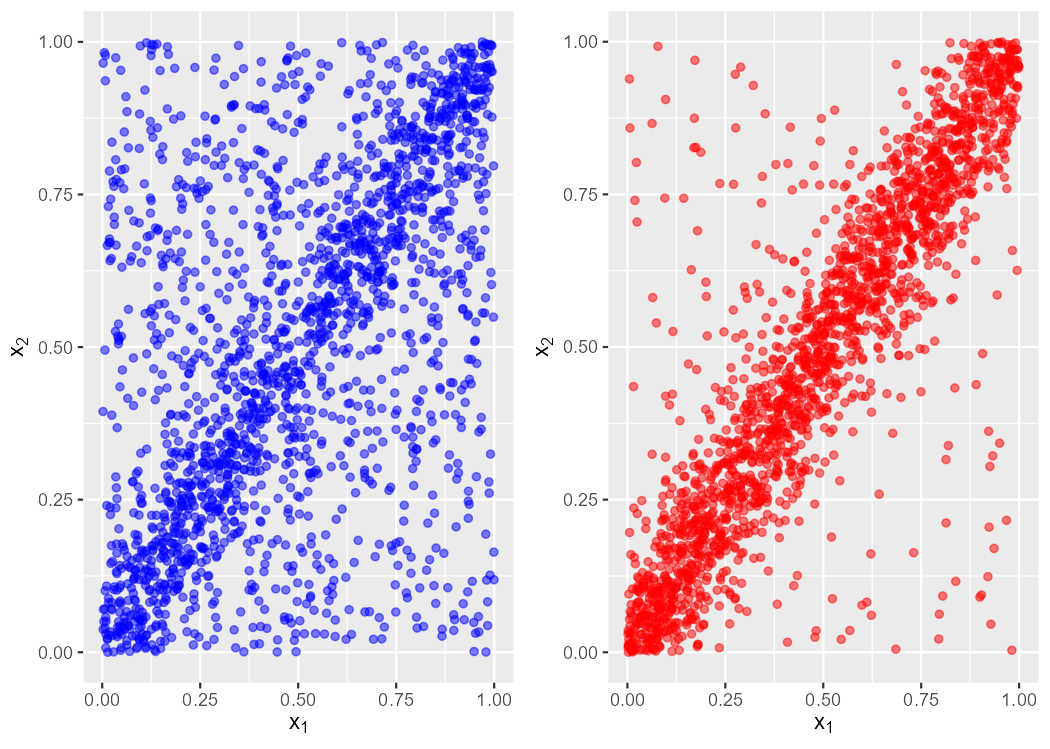}
\caption{Case Study \#10: UniformFrankD2 (equal marginals).}
\end{figure}
\newpage
\subsubsection{Case Study 11:  Pareto Simplex copulas}
\begin{itemize}
\item
x: observations from a Pareto Simplex copula with parameter 1.
\item
y: observations from a Pareto Simplex copula with parameter $\theta$.
\end{itemize}
 
\begin{figure}[!htbp]
\centering
\includegraphics[width=4in]{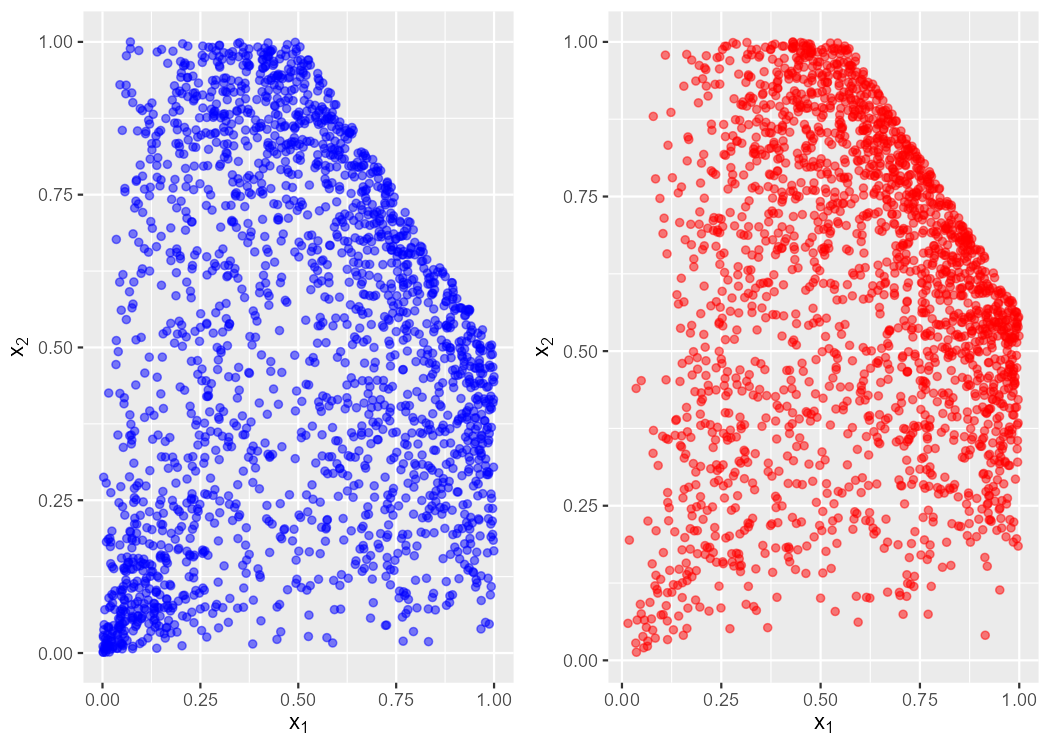}
\caption{Case Study \#11: ParetoSimplexD2 (equal marginals).}
\end{figure}
\newpage
\subsubsection{Case Study 12: KhoudrajiClayton copulas}
\begin{itemize}
\item
x: observations from KhoudrajiClayton with shape parameters = $(0.2, 0.95)$.
\item
y: observations from KhoudrajiClayton with shape parameters = $(\alpha, 0.95)$.
\end{itemize}
 
\begin{figure}[!htbp]
\centering
\includegraphics[width=4in]{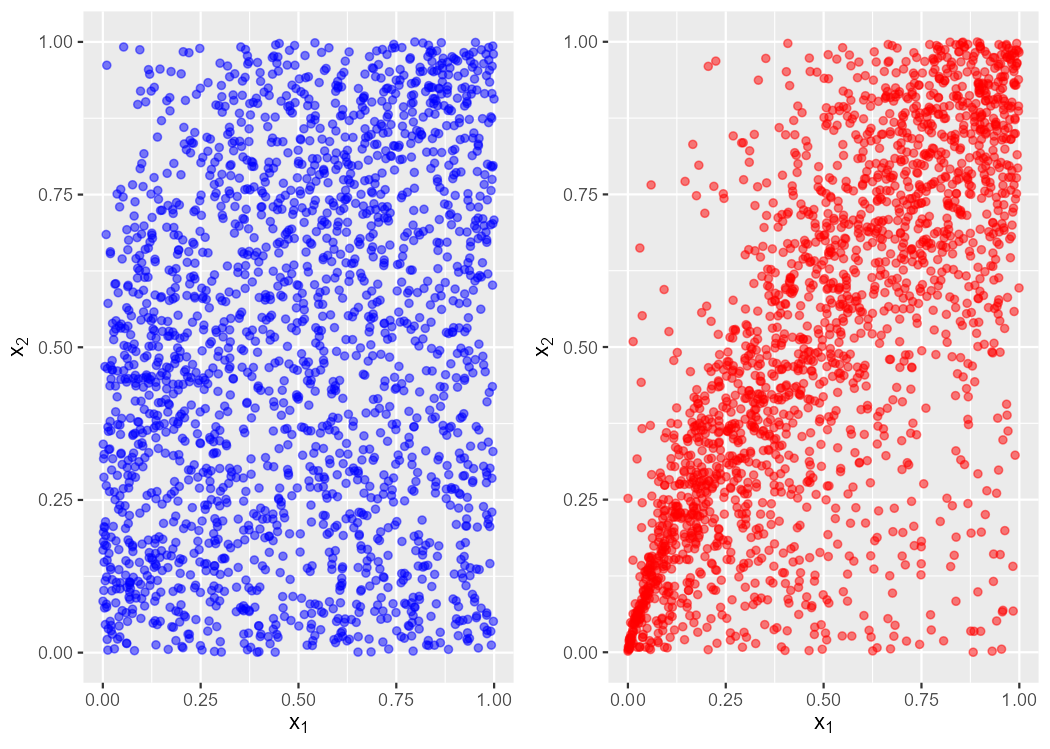}
\caption{Case Study \#12: KhoudrajiClaytonD2 (equal marginals).}
\end{figure}
\newpage
\subsubsection{Case Study 13: Mixture of Normal and Uniform copulas}
\begin{itemize}
\item
x: observations from a 50-50 mixture of Normal and Uniform copulas.
\item
y: observations from a $(\alpha, 1-\alpha)$ mixture of Normal and Uniform copulas.
\end{itemize}
 
\begin{figure}[!htbp]
\centering
\includegraphics[width=4in]{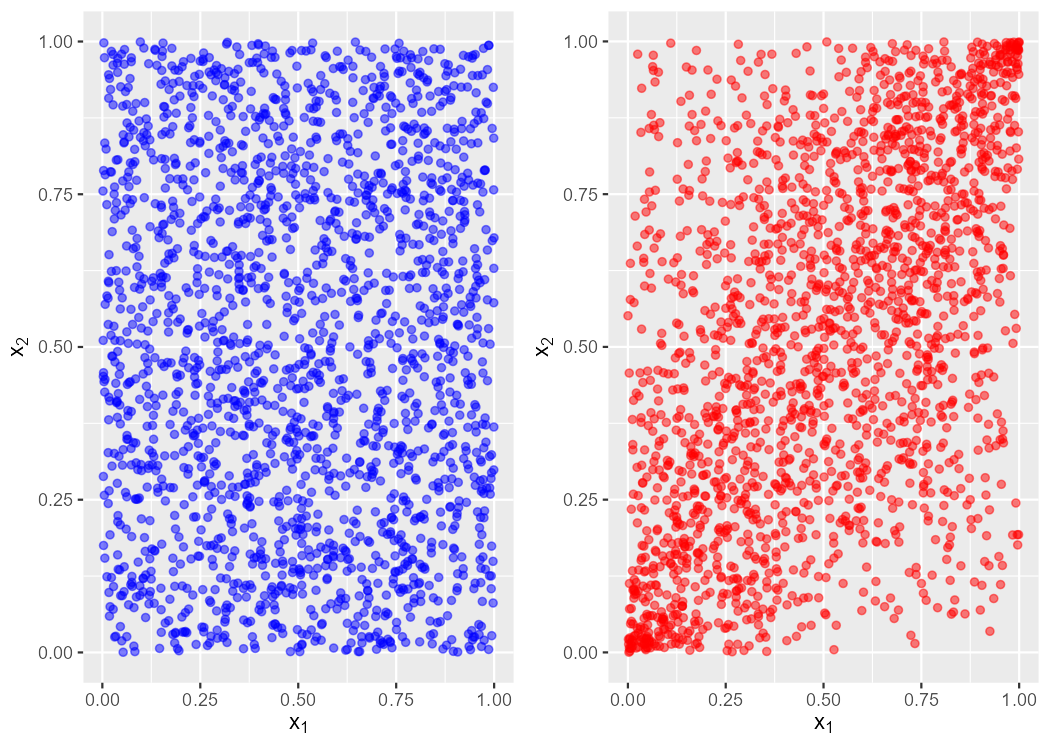}
\caption{Case Study \#13: NormalUniformD2 (equal marginals).}
\end{figure}
\newpage
\subsubsection{Case Study 14: Joe copulas}
\begin{itemize}
\item
x: observations from a Joe copula with parameter 1.
\item
y: observations from a Joe copula with parameter  $\alpha$.
\end{itemize}
 
\begin{figure}[!htbp]
\centering
\includegraphics[width=4in]{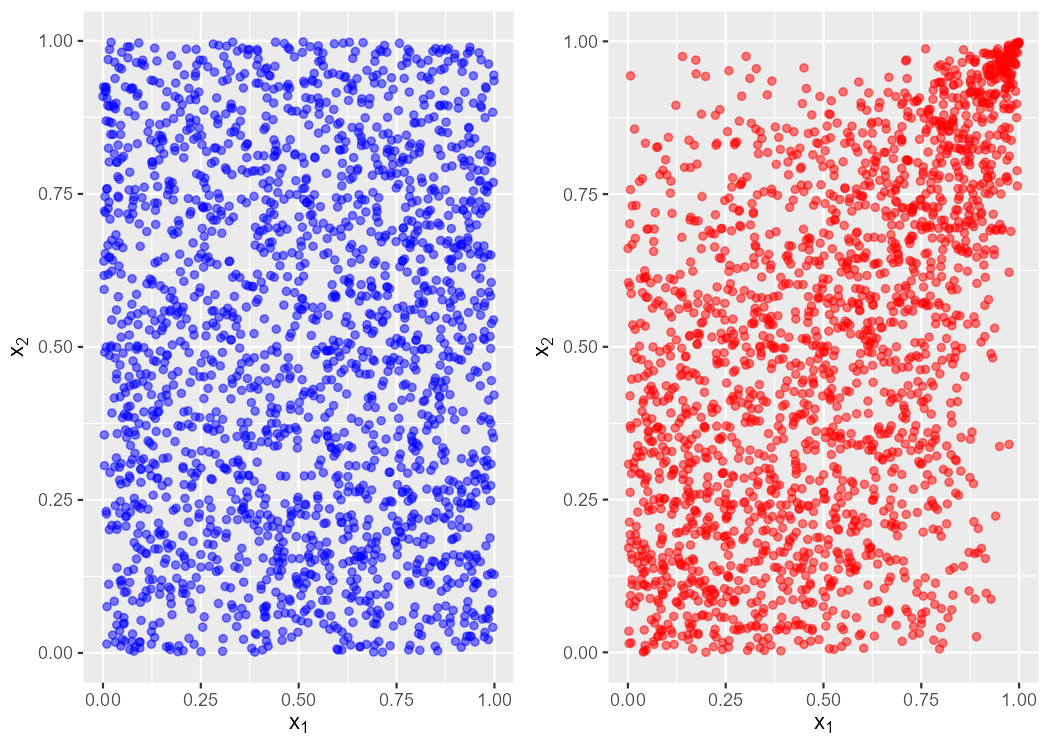}
\caption{Case Study \#14: JoeD2 (equal marginals).}
\end{figure}
\newpage
\subsubsection{Case Study 15: CLEO Dalitz Plot with Diagonal Stripe}
\begin{itemize}
\item
x: observations from a Cleo Dalitz plot.
\item
y: observations from a Cleo Dalitz plot with diagonal stripe.
\end{itemize}
 
\begin{figure}[!htbp]
\centering
\includegraphics[width=4in]{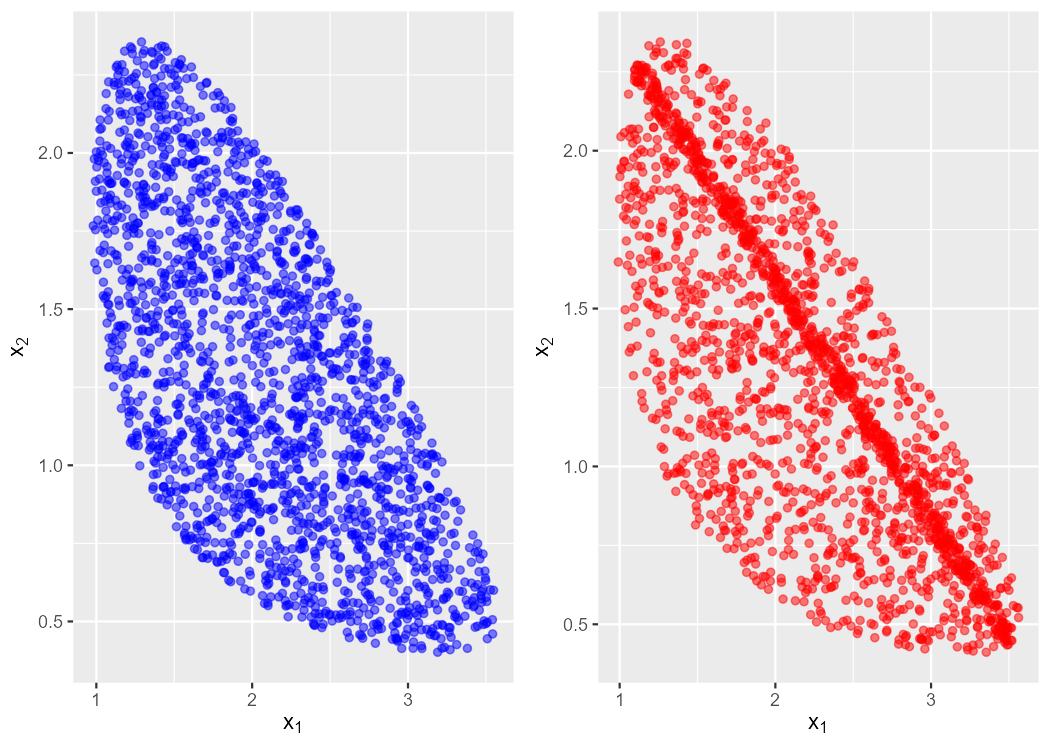}
\caption{Case Study \#15: DalitzCleoD2 (equal marginals).}
\end{figure}
\newpage
\subsubsection{Case Study 16: PDG Dalitz Plot}
\begin{itemize}
\item
x: observations from a PDG Dalitz plot.
\item
y: observations from a PDG Dalitz plot with diagonal stripe.
\end{itemize}
 
\begin{figure}[!htbp]
\centering
\includegraphics[width=4in]{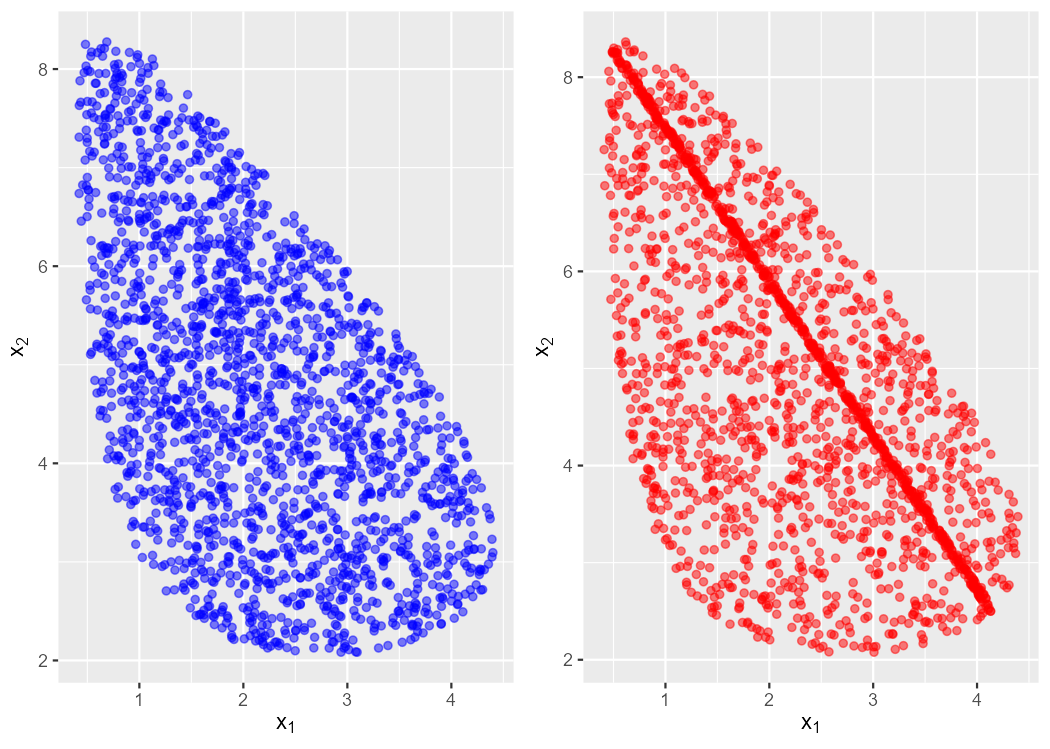}
\caption{Case Study \#16: DalitzPDGD2 (equal marginals).}
\end{figure}
\newpage
\subsubsection{Case Study 17: Babar Dalitz Plot}
\begin{itemize}
\item
x: observations from a Babar Dalitz plot.
\item
y: observations from a Babar Dalitz plot with diagonal stripe.
\end{itemize}
 
\begin{figure}[!htbp]
\centering
\includegraphics[width=4in]{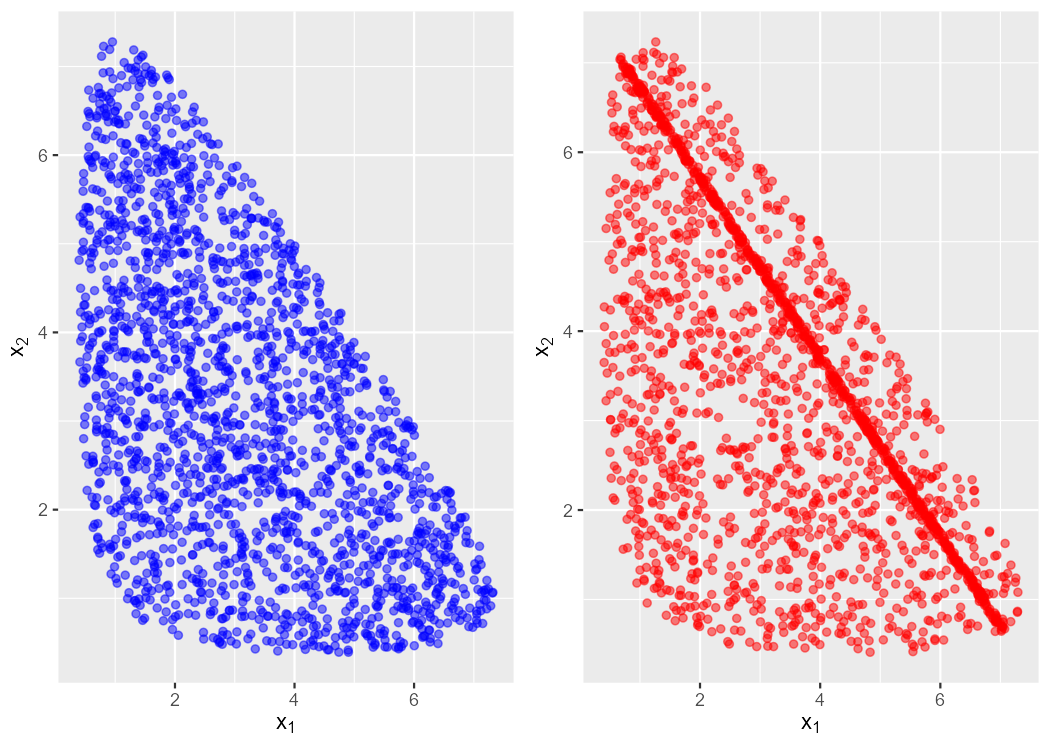}
\caption{Case Study \#17: DalitzBabarD2 (equal marginals).}
\end{figure}
\newpage
\subsubsection{Case Study 18: Multivariate normal distribution with shift}
\begin{itemize}
\item
x: observations from a bivariate normal distribution with mean vector $(0,0)$ and variance-covariance matrix the identity.
\item
y:  observations from a bivariate normal distribution with mean vector $(0,\mu)$ and variance-covariance matrix the identity.
\end{itemize}
 
\begin{figure}[!htbp]
\centering
\includegraphics[width=4in]{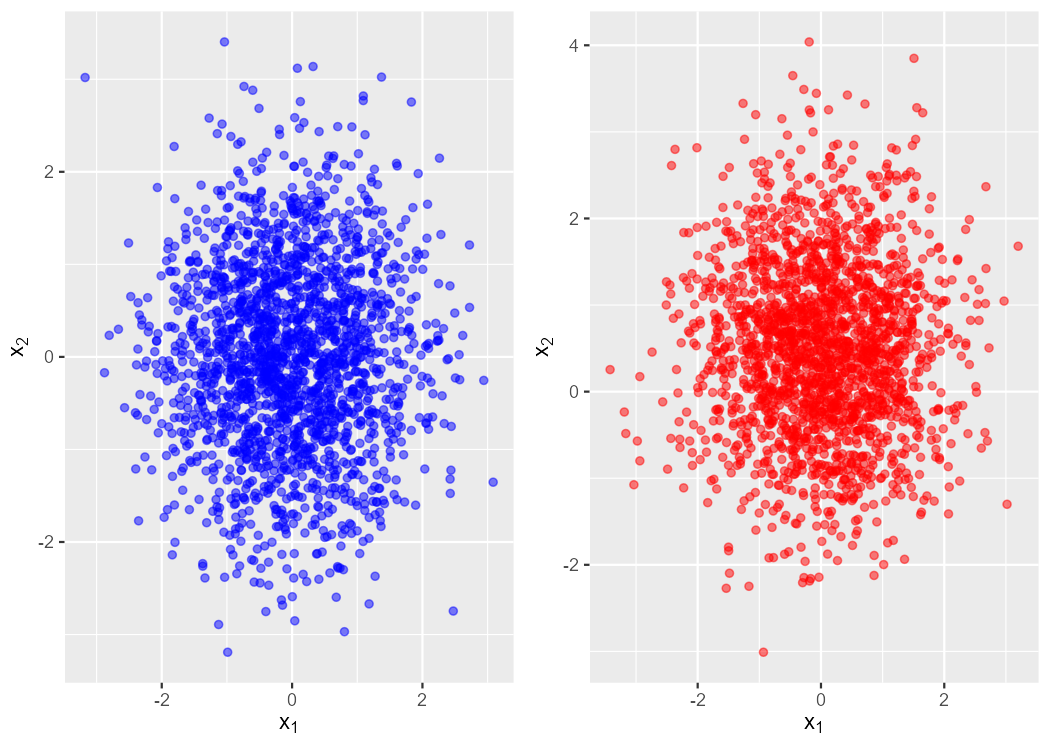}
\caption{Case Study \#18: NormalShiftM (unequal marginals).}
\end{figure}
\newpage
\subsubsection{Case Study 19: Multivariate normal distribution with stretch}
\begin{itemize}
\item
x: observations from a bivariate normal distribution with mean vector $(0,0)$ and variance-covariance matrix the identity.
\item
y:  observations from a bivariate normal distribution with mean vector $(0,0)$ and variance-covariance matrix matrix $\begin{bmatrix} 	1	&  1 \\ 	1	&  s  \end{bmatrix}$.
\end{itemize}
 
\begin{figure}[!htbp]
\centering
\includegraphics[width=4in]{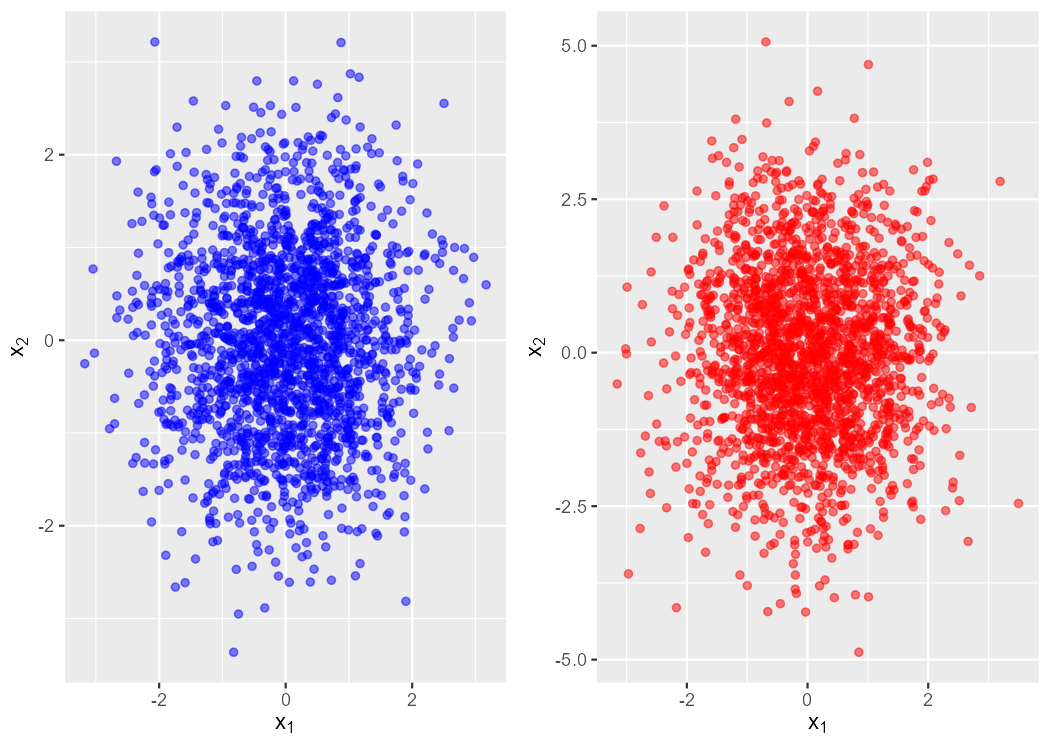}
\caption{Case Study \#19: NormalStretchM (unequal marginals).}
\end{figure}
\newpage
\subsubsection{Case Study 20: Uniform distribution with rotation}
\begin{itemize}
\item
x: two independent uniform random variables.
\item
y: observations from independent rotated uniform distributions.
\end{itemize}
 
\begin{figure}[!htbp]
\centering
\includegraphics[width=4in]{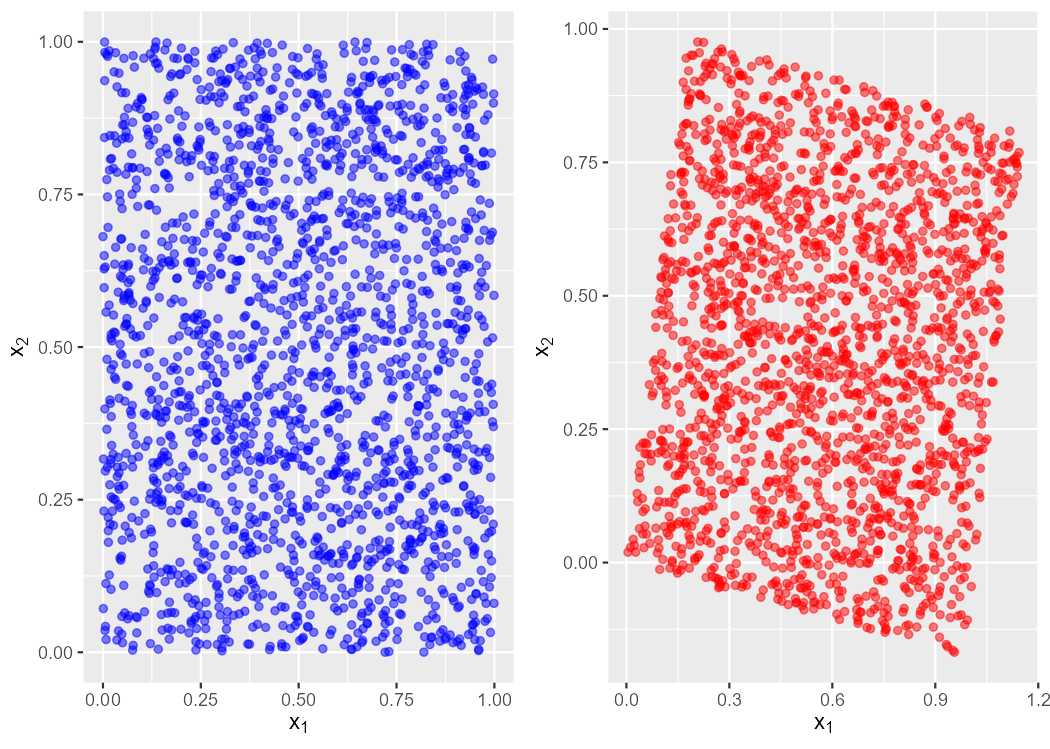}
\caption{Case Study \#20: UniformRotateM (unequal marginals).}
\end{figure}
\newpage
\subsubsection{Case Study 21: Uniform vs Beta distributions}
\begin{itemize}
\item
x: two independent uniform random variables.
\item
y: observations from independent Beta distributions.
\end{itemize}
 
\begin{figure}[!htbp]
\centering
\includegraphics[width=4in]{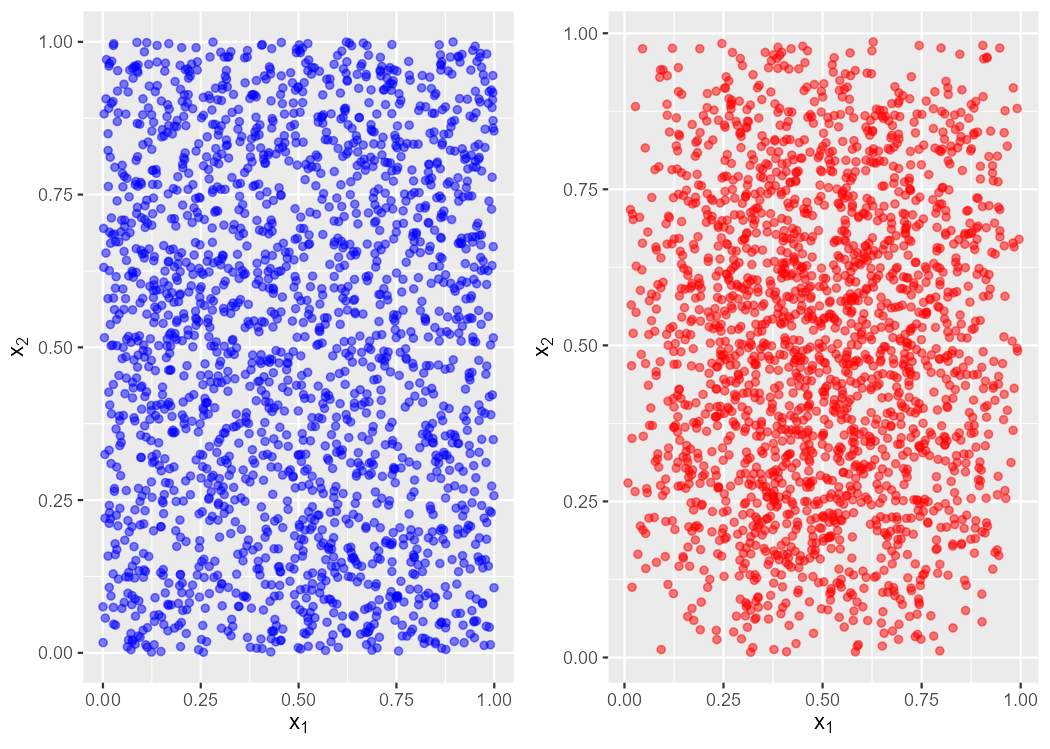}
\caption{Case Study \#21: UniformBetaM (unequal marginals).}
\end{figure}
\newpage
\subsubsection{Case Study 22: Truncated Exponential distributions}
\begin{itemize}
\item
x: observations from independent exponential distributions with rates 1, truncated to the interval $[0,2]$.
\item
y:  observations from independent exponential distributions with rates $\lambda$, truncated to the interval $[0,2]$.
\end{itemize}
 
\begin{figure}[!htbp]
\centering
\includegraphics[width=4in]{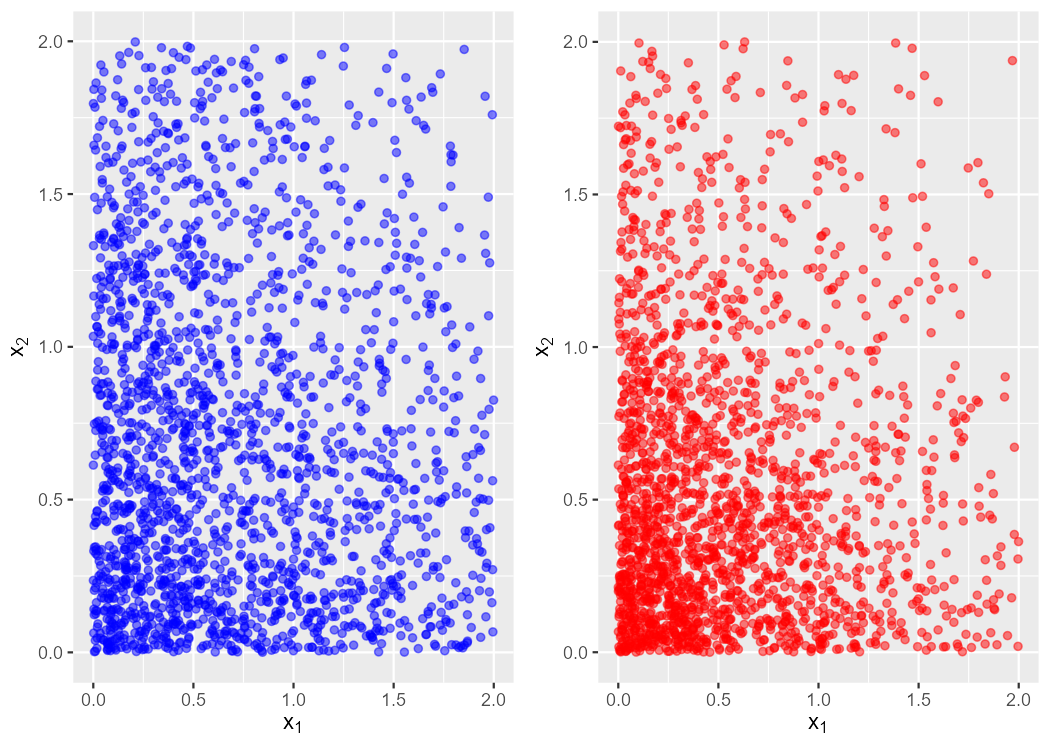}
\caption{Case Study \#22: TruncExponentialM (unequal marginals).}
\end{figure}
\newpage
\subsubsection{Case Study 23: CLEO Dalitz Plot with additional features}
\begin{itemize}
\item
x: observations from a Cleo Dalitz plot.
\item
y: observations from a Cleo Dalitz plot with additional features.
\end{itemize}
 
\begin{figure}[!htbp]
\centering
\includegraphics[width=4in]{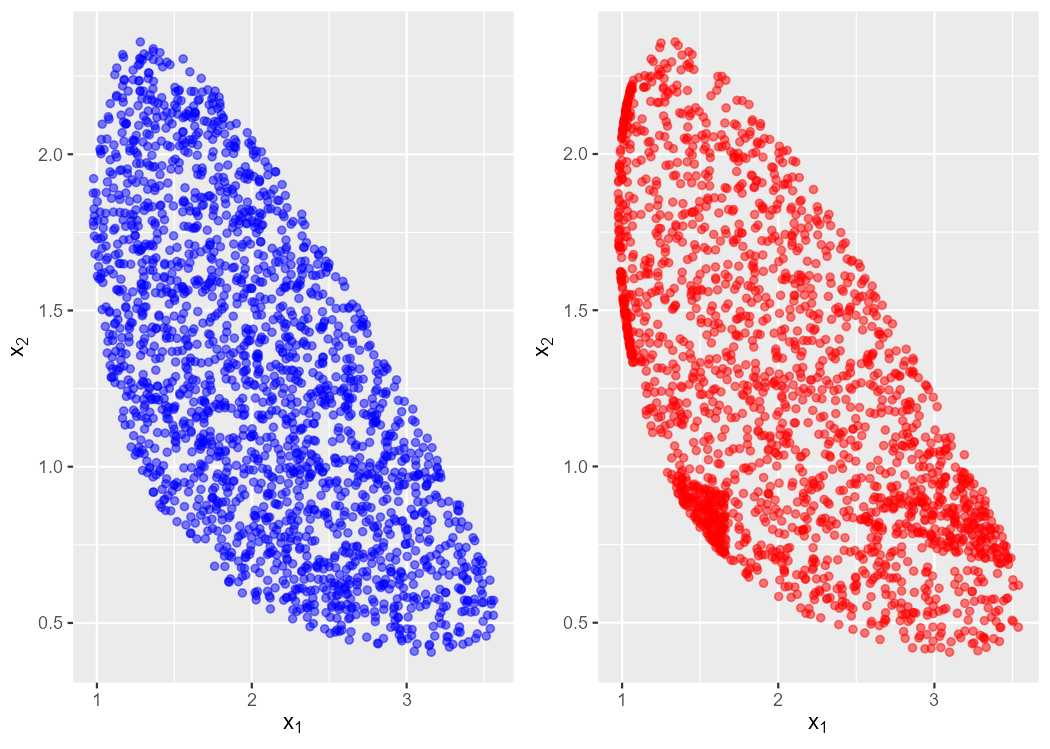}
\caption{Case Study \#23: DalitzCleoM (unequal marginals).}
\end{figure}
\newpage
\subsubsection{Case Study 24: PDG Dalitz Plot additional features}
\begin{itemize}
\item
x: observations from a PDG Dalitz plot.
\item
y: observations from a PDG Dalitz plot with additional features.
\end{itemize}
 
\begin{figure}[!htbp]
\centering
\includegraphics[width=4in]{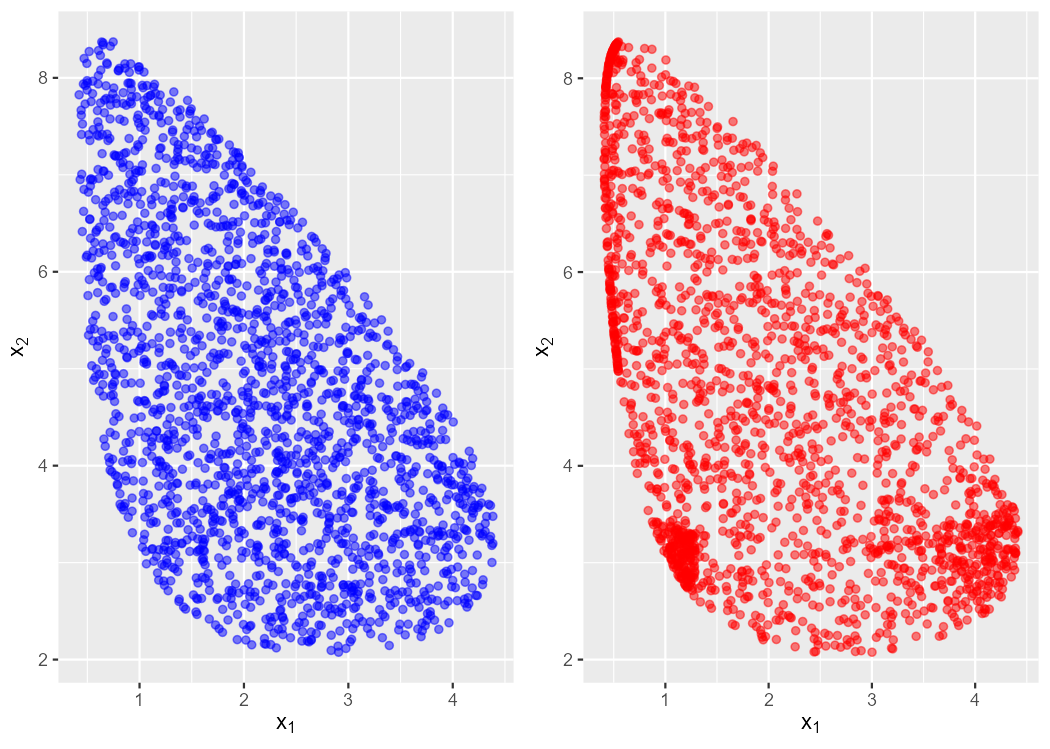}
\caption{Case Study \#24: DalitzPDGM (unequal marginals).}
\end{figure}
\newpage
\subsubsection{Case Study 25: Babar Dalitz Plot additional features}
\begin{itemize}
\item
x: observations from a Babar Dalitz plot.
\item
y: observations from a Babar Dalitz plot with additional features.
\end{itemize}
 
\begin{figure}[!htbp]
\centering
\includegraphics[width=4in]{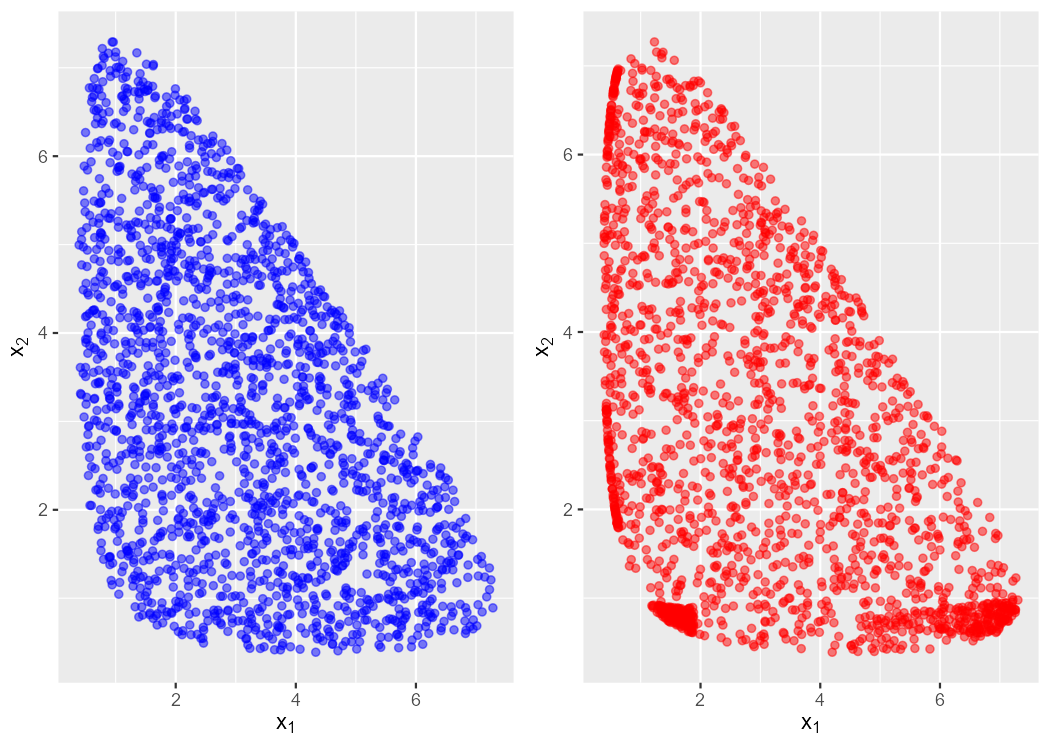}
\caption{Case Study \#25: DalitzBabarM (unequal marginals).}
\end{figure}
\newpage
\subsubsection{Case Study 26: Frank coupola with exponential marginals}
\begin{itemize}
\item
x: observations from a Frank coupola with exponential rate 1 marginals.
\item
y: observations from a Frank coupola with exponential rate $\lambda$ marginals.
\end{itemize}
 
\begin{figure}[!htbp]
\centering
\includegraphics[width=4in]{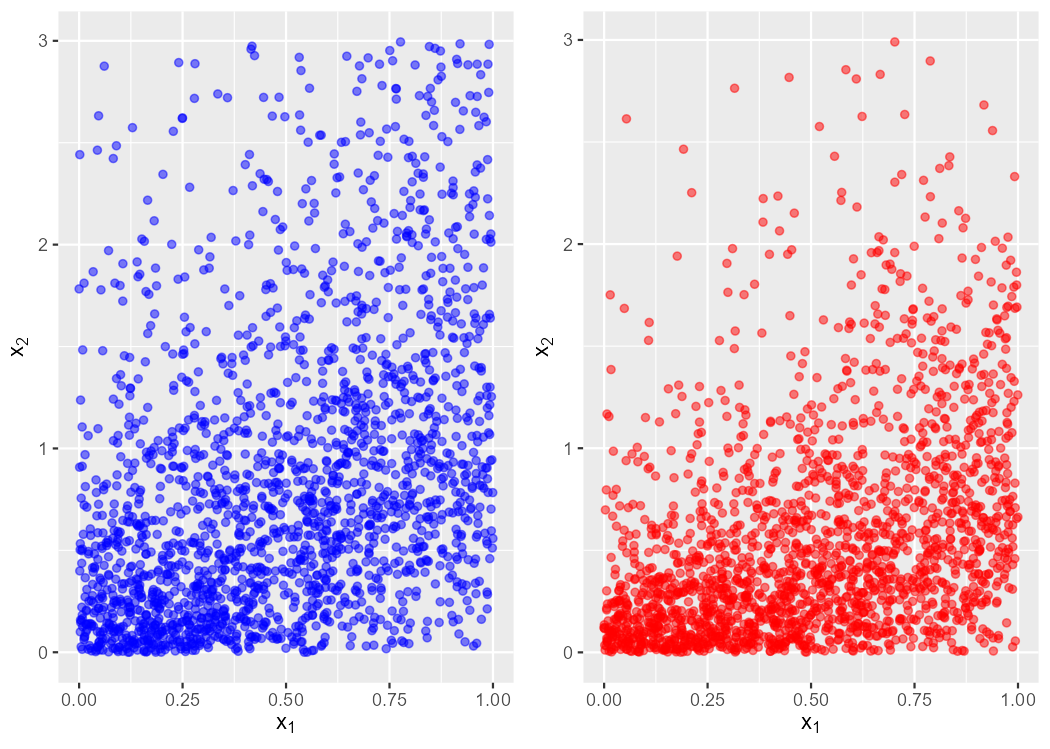}
\caption{Case Study \#26: FrankExponentialM (unequal marginals).}
\end{figure}
\newpage
\subsubsection{Case Study 27: Frank coupola with linear marginals}
\begin{itemize}
\item
x: observations from a Frank coupola with linear marginals.
\item
y: observations from a Frank coupola with linear marginals with a different slope.
\end{itemize}
 
\begin{figure}[!htbp]
\centering
\includegraphics[width=4in]{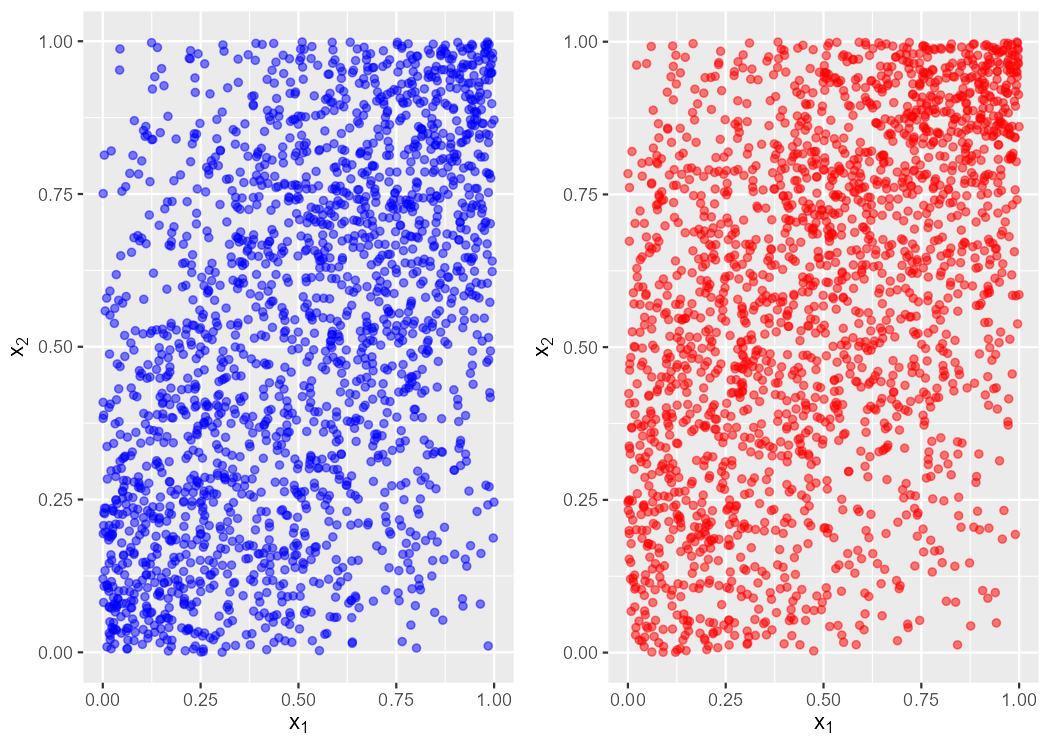}
\caption{Case Study \#27: FrankLinearM (unequal marginals).}
\end{figure}
\newpage
\subsubsection{Case Study 28: Frank coupola with normal tail marginals}
\begin{itemize}
\item
x: observations from a Frank coupola with standard normal tail marginals.
\item
y: observations from a Frank coupola with normal tail marginals with standard deviation $\sigma$.
\end{itemize}
 
\begin{figure}[!htbp]
\centering
\includegraphics[width=4in]{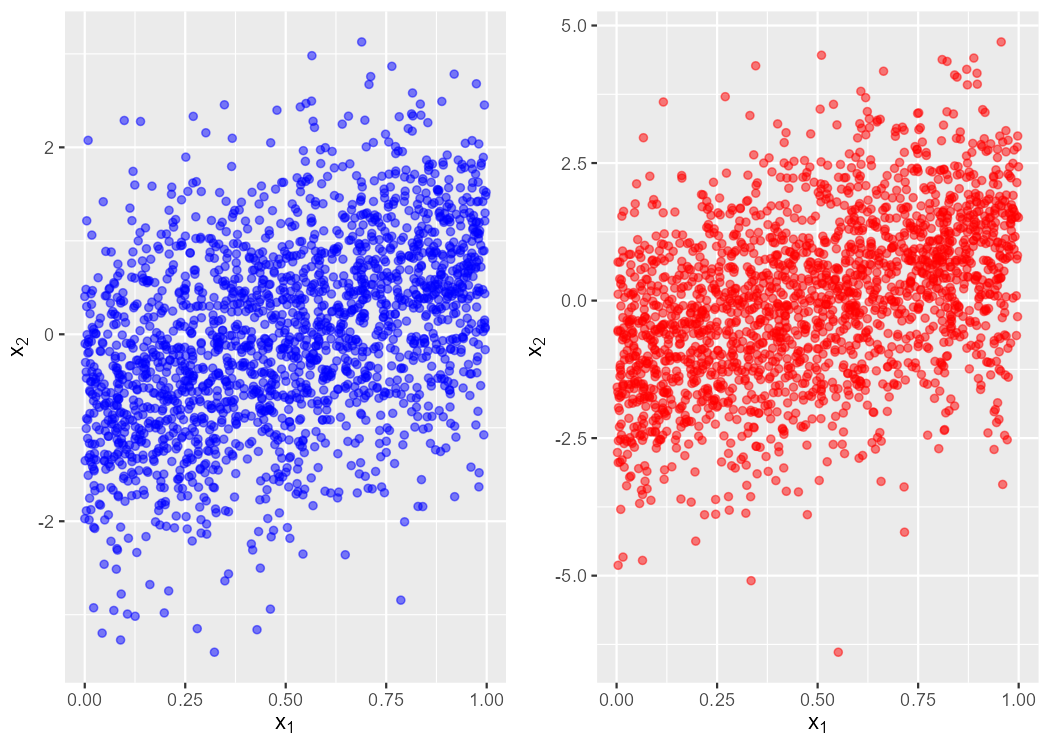}
\caption{Case Study \#28: FrankNormalM (unequal marginals).}
\end{figure}
\newpage
\subsubsection{Case Study 29: Clayton coupola with exponential marginals}
\begin{itemize}
\item
x: observations from a Clayton coupola with exponential rate 1 marginals.
\item
y: observations from a Clayton coupola with exponential rate $\lambda$ marginals.
\end{itemize}
 
\begin{figure}[!htbp]
\centering
\includegraphics[width=4in]{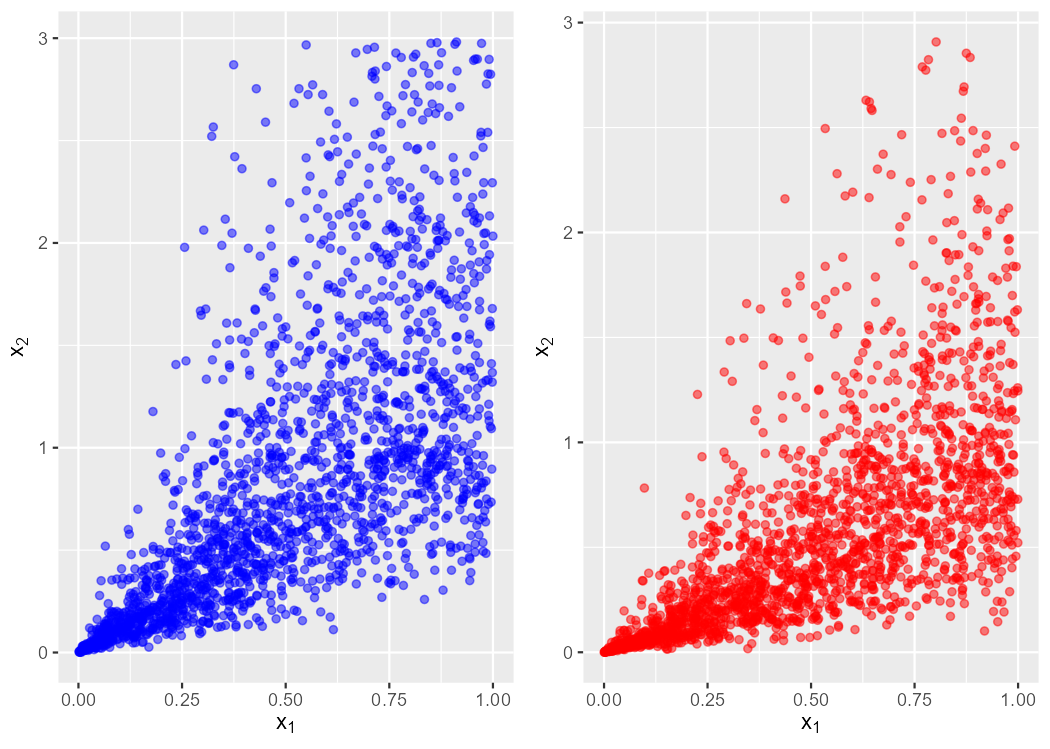}
\caption{Case Study \#29: ClaytonExponentialM (unequal marginals).}
\end{figure}
\newpage
\subsubsection{Case Study 30: Clayton coupola with linear marginals}
\begin{itemize}
\item
x: observations from a Clayton coupola with linear marginals.
\item
y: observations from a Clayton coupola with linear marginals with a different slope.
\end{itemize}
 
\begin{figure}[!htbp]
\centering
\includegraphics[width=4in]{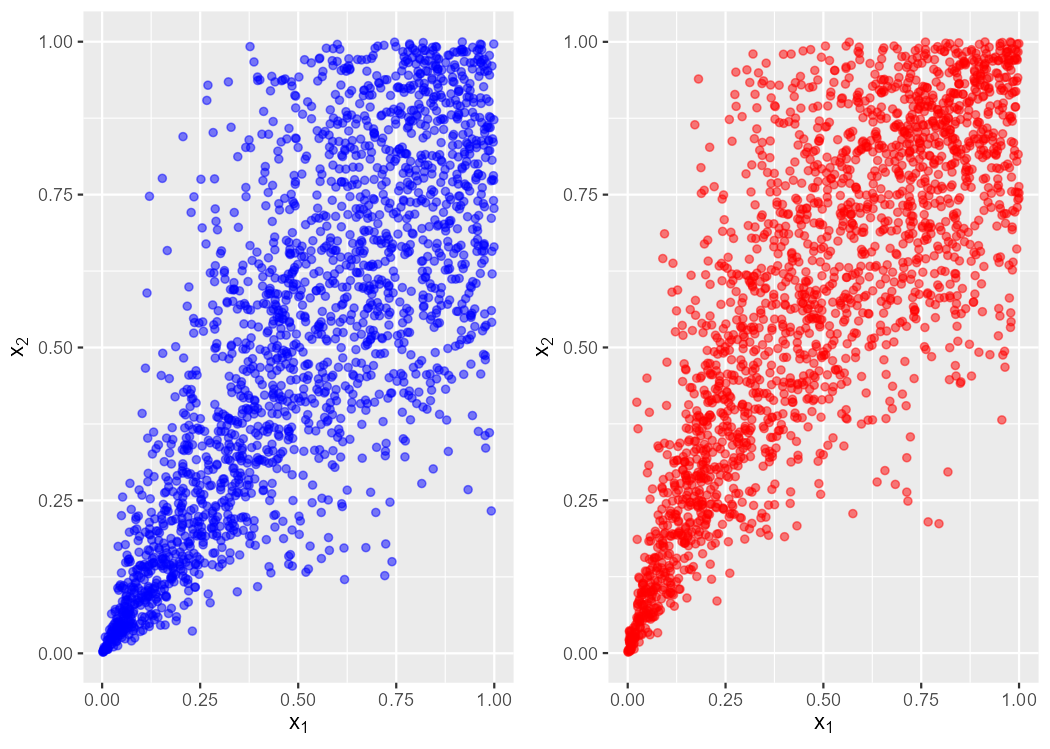}
\caption{Case Study \#30: ClaytonLinearM (unequal marginals).}
\end{figure}
\newpage
\subsubsection{Case Study 31: Clayton coupola with normal tail marginals}
\begin{itemize}
\item
x: observations from a Clayton coupola with standard normal tail marginals.
\item
y: observations from a Clayton coupola with normal tail marginals with standard deviation $\sigma$.
\end{itemize}
 
\begin{figure}[!htbp]
\centering
\includegraphics[width=4in]{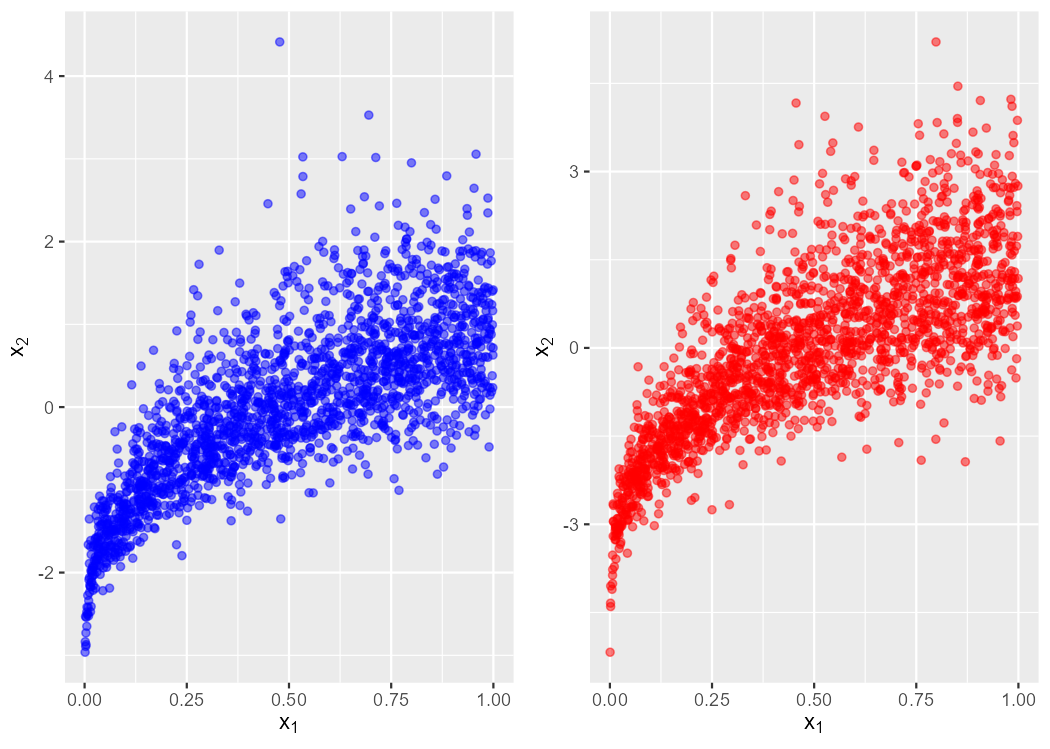}
\caption{Case Study \#31: ClaytonNormalM (unequal marginals).}
\end{figure}
\newpage
\subsubsection{Case Study 32: Galambos coupola with exponential marginals}
\begin{itemize}
\item
x: observations from a Galambos coupola with exponential rate 1 marginals.
\item
y: observations from a Galambos coupola with exponential rate $\lambda$ marginals.
\end{itemize}
 
\begin{figure}[!htbp]
\centering
\includegraphics[width=4in]{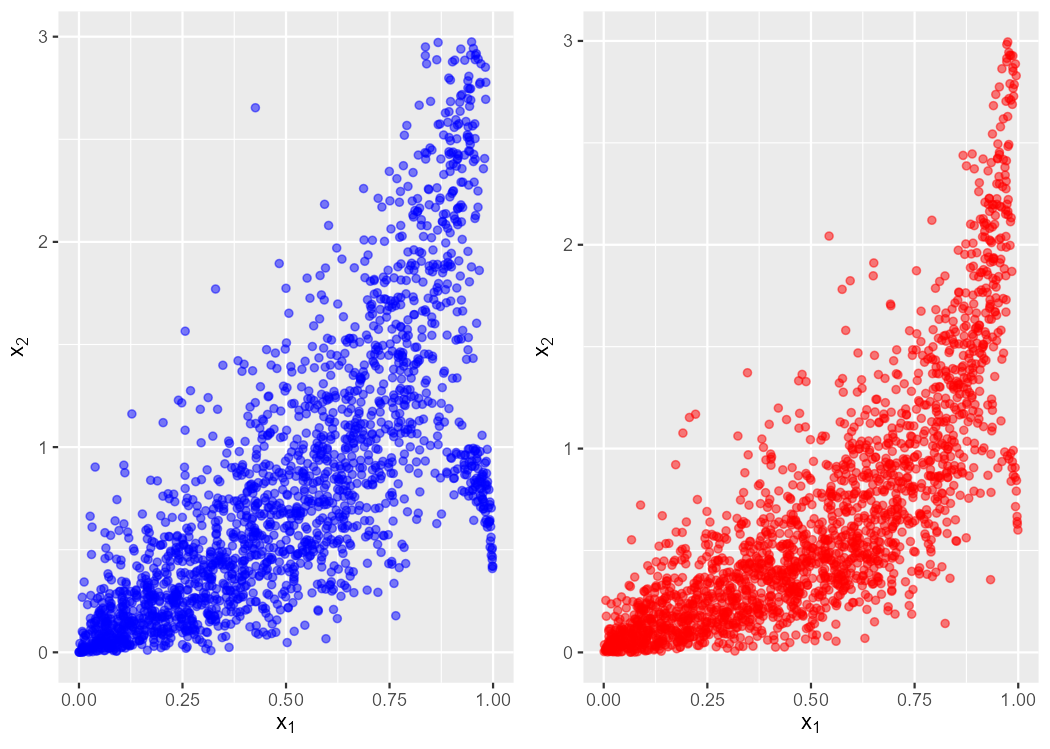}
\caption{Case Study \#32: GalambosExponentialM (unequal marginals).}
\end{figure}
\newpage
\subsubsection{Case Study 33: Galambos coupola with linear marginals}
\begin{itemize}
\item
x: observations from a Galambos coupola with linear marginals.
\item
y: observations from a Galambos coupola with linear marginals with a different slope.
\end{itemize}
 
\begin{figure}[!htbp]
\centering
\includegraphics[width=4in]{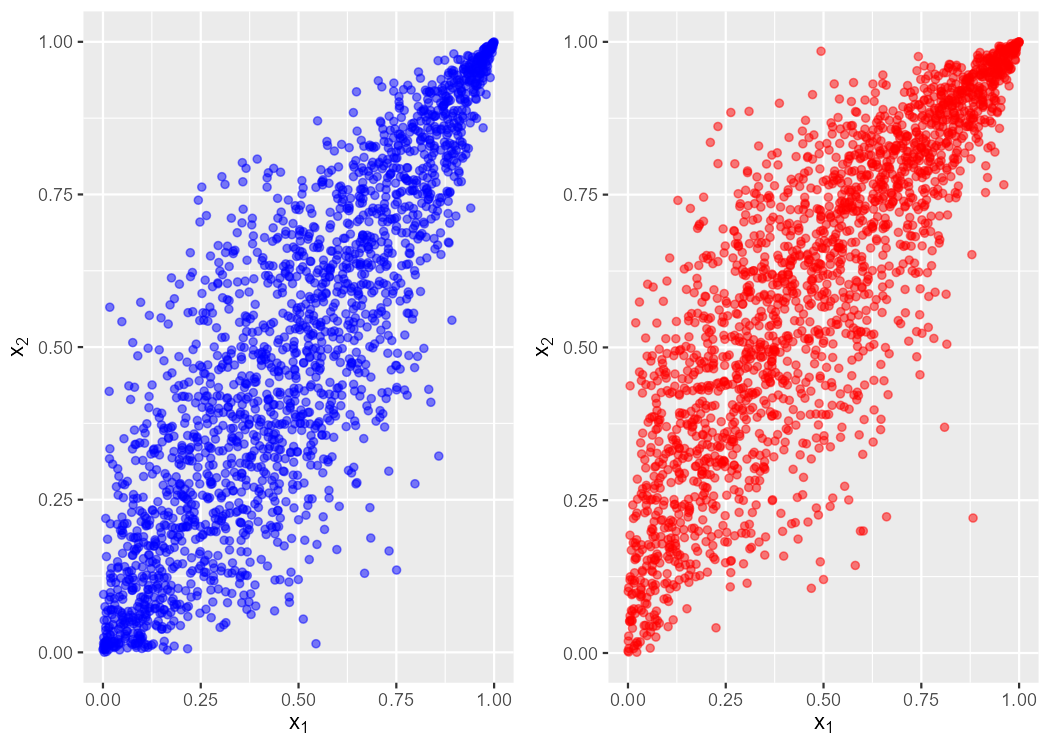}
\caption{Case Study \#33: GalambosLinearM (unequal marginals).}
\end{figure}
\newpage
\subsubsection{Case Study 34: Galambos coupola with normal tail marginals}
\begin{itemize}
\item
x: observations from a Galambos coupola with standard normal tail marginals.
\item
y: observations from a Galambos coupola with normal tail marginals with standard deviation $\sigma$.
\end{itemize}
 
\begin{figure}[!htbp]
\centering
\includegraphics[width=4in]{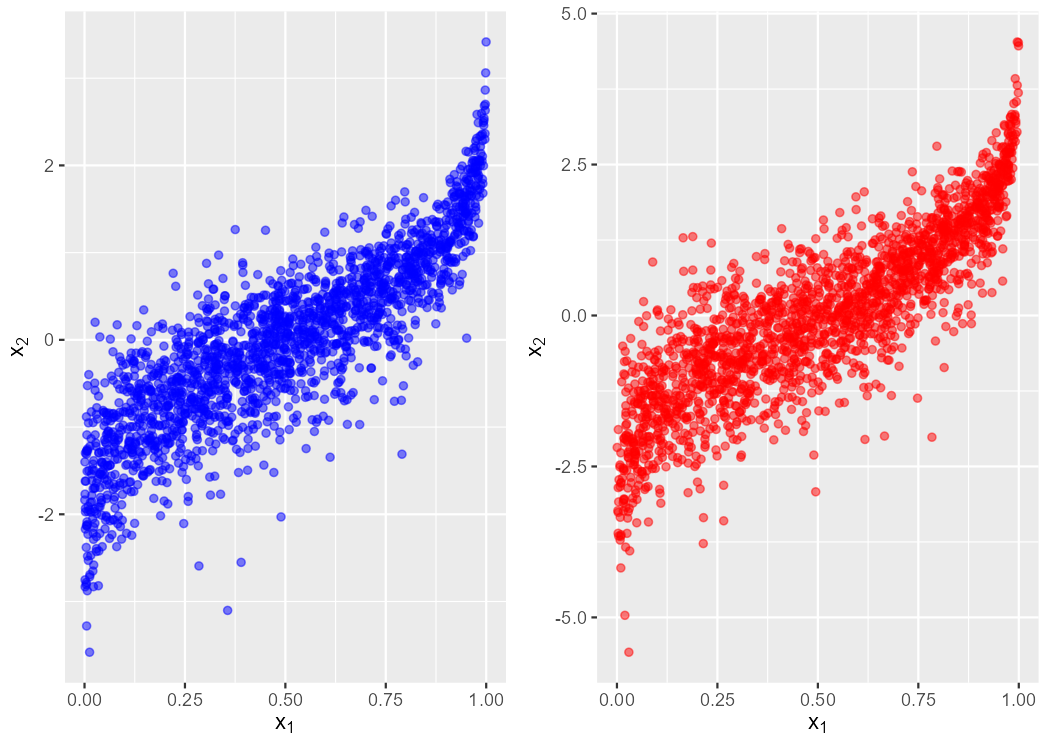}
\caption{Case Study \#34: GalambosNormalM (unequal marginals).}
\end{figure}
\newpage

\subsection{Five dimensional data}

Here we have the following case studies:

\emph{Case studies with equal marginals:}

\begin{itemize}
\item
  NormalD5:  multivariate normal distribution with equal marginals.
\item
  tD5:  multivariate t distribution with 5 degrees of freedom and equal marginals.                 
\item
  FrankD5: Frank cupola.
\item
  ClaytonD5: Clayton cupola.
\item
  GumbelD5: Gumbel copula.            
\item
  JoeD5: Joe cupola.
\item 
  UniformFrankD5: mixture of uniform and Frank cupola.
\item
  FrankClaytonD5: mixture of Frank and Clayton cupolas.
\item	
  FrankJoeD5: mixture of Frank and Joe cupolas.           
\end{itemize}	

\emph{Case studies with unequal marginals:}

\begin{itemize}
\item
	UniformExponentialM5: Exponential distributions.
\item
  FrankExponentialM5: Frank cupola with exponential marginals.
\item
  FrankLinearM5: Frank cupola with linear marginals.
\item
  FrankNormalM5: Frank cupola with linear marginals.
\item
  ClaytonExponentialM5: Clayton cupola with exponential marginals.
\item
  ClaytonLinearM5: Clayton cupola with linear marginals.
\item
  ClaytonNormalM5: Clayton cupola with linear marginals.
\end{itemize}

\section{Appendix 1: full results}

\subsection{Type I error probabilities and power estimates}

\subsubsection{Continuous data - type I error}

% latex table generated in R 4.5.1 by xtable 1.8-4 package
% Tue Jul 22 10:01:20 2025
\begin{table}[ht]
\centering
\begin{tabular}{rrrrrrrrrrr}
  \hline
 & KS & K & CvM & AD & NN1 & NN5 & AZ & BF & BG & FR \\ 
  \hline
NormalD2 & 5.90 & 5.10 & 6.30 & 5.50 & 4.80 & 4.30 & 5.40 & 4.90 & 4.80 & 5.50 \\ 
  tD2 & 3.90 & 4.90 & 4.00 & 3.50 & 4.20 & 3.70 & 4.30 & 3.80 & 3.60 & 4.20 \\ 
  UniformMixtureD2 & 5.50 & 3.70 & 5.30 & 6.00 & 5.70 & 3.50 & 4.20 & 4.20 & 4.40 & 5.10 \\ 
  FrankD2 & 4.80 & 3.70 & 5.30 & 5.70 & 5.00 & 4.40 & 4.30 & 4.60 & 6.10 & 5.30 \\ 
  ClaytonD2 & 5.80 & 5.90 & 5.90 & 6.40 & 4.60 & 4.70 & 6.00 & 5.50 & 5.90 & 5.40 \\ 
  GumbelD2 & 4.40 & 3.50 & 5.30 & 5.30 & 4.40 & 3.00 & 3.50 & 4.10 & 4.30 & 4.30 \\ 
  GalambosD2 & 4.40 & 4.10 & 4.30 & 3.20 & 3.80 & 5.80 & 3.60 & 2.80 & 3.70 & 5.30 \\ 
  HuslerReissD2 & 4.90 & 5.40 & 5.20 & 4.60 & 5.90 & 5.80 & 5.30 & 6.70 & 5.10 & 5.80 \\ 
  ClaytonGumbelD2 & 5.20 & 5.60 & 4.10 & 4.80 & 4.80 & 6.90 & 5.90 & 5.80 & 5.00 & 5.60 \\ 
  UniformFrankD2 & 7.60 & 6.70 & 5.90 & 6.50 & 5.90 & 5.50 & 5.70 & 5.90 & 4.10 & 5.30 \\ 
  ParetoSimplexD2 & 5.80 & 4.30 & 5.80 & 6.90 & 3.30 & 4.60 & 6.30 & 7.00 & 6.30 & 5.40 \\ 
  KhoudrajiClaytonD2 & 4.20 & 4.70 & 5.10 & 4.30 & 4.60 & 5.90 & 4.80 & 4.30 & 7.60 & 4.80 \\ 
  NormalUniformD2 & 4.60 & 2.70 & 5.90 & 5.90 & 4.70 & 4.90 & 4.60 & 5.80 & 6.40 & 5.20 \\ 
  JoeD2 & 5.40 & 4.90 & 6.50 & 5.50 & 4.80 & 3.90 & 6.20 & 6.40 & 5.00 & 4.60 \\ 
  DalitzCleoD2 & 4.80 & 5.80 & 7.20 & 6.60 & 3.50 & 5.70 & 7.50 & 6.50 & 4.80 & 4.70 \\ 
  DalitzPDGD2 & 4.20 & 6.30 & 6.00 & 5.90 & 5.80 & 5.80 & 6.30 & 6.10 & 5.50 & 3.80 \\ 
  DalitzBabarD2 & 3.50 & 6.00 & 4.00 & 4.30 & 5.10 & 3.20 & 5.30 & 5.30 & 5.20 & 5.30 \\ 
  NormalShiftM & 5.50 & 5.30 & 4.90 & 5.40 & 5.60 & 3.90 & 4.00 & 4.10 & 5.00 & 5.60 \\ 
  NormalStretchM & 5.00 & 4.90 & 4.90 & 5.20 & 4.70 & 4.90 & 5.10 & 5.10 & 5.70 & 5.20 \\ 
  UniformRotateM & 4.00 & 4.00 & 4.40 & 4.60 & 3.40 & 4.60 & 4.60 & 5.30 & 5.50 & 4.80 \\ 
  UniformBetaM & 5.40 & 5.10 & 4.10 & 4.40 & 7.30 & 6.10 & 4.30 & 4.30 & 4.20 & 5.80 \\ 
  TruncExponentialM & 4.20 & 5.30 & 4.90 & 4.40 & 5.20 & 5.20 & 4.80 & 4.30 & 5.50 & 5.10 \\ 
  DalitzCleoM & 5.20 & 6.10 & 5.20 & 4.60 & 6.50 & 5.90 & 4.60 & 4.40 & 4.90 & 5.00 \\ 
  DalitzPDGM & 4.30 & 4.90 & 5.30 & 5.40 & 8.60 & 6.90 & 5.50 & 4.80 & 5.30 & 4.80 \\ 
  DalitzBabarM & 6.20 & 6.10 & 4.70 & 5.00 & 3.90 & 4.60 & 4.40 & 4.60 & 7.30 & 3.90 \\ 
  NormalD5 & 4.10 & 3.50 & 5.50 & 5.90 & 5.10 & 5.00 & 7.20 & 6.20 & 5.60 & 5.00 \\ 
  tD5 & 3.20 & 2.20 & 4.90 & 3.80 & 4.70 & 5.00 & 4.20 & 4.60 & 5.00 & 5.00 \\ 
  FrankD5 & 6.70 & 6.20 & 5.30 & 4.00 & 4.30 & 3.80 & 4.90 & 4.30 & 4.20 & 5.70 \\ 
  ClaytonD5 & 4.30 & 5.10 & 3.80 & 3.10 & 4.60 & 5.40 & 3.80 & 3.70 & 4.90 & 3.90 \\ 
  GumbelD5 & 4.90 & 6.30 & 6.20 & 5.50 & 5.00 & 6.00 & 4.20 & 4.30 & 5.20 & 5.10 \\ 
  JoeD5 & 4.10 & 3.70 & 6.40 & 5.30 & 4.80 & 5.10 & 5.20 & 5.10 & 5.30 & 5.10 \\ 
  FrankExponentialM & 3.70 & 4.60 & 5.50 & 5.30 & 6.80 & 6.00 & 5.60 & 5.90 & 6.30 & 5.30 \\ 
  FrankLinearM & 7.10 & 6.20 & 7.70 & 7.20 & 4.40 & 4.10 & 5.60 & 5.80 & 6.10 & 5.80 \\ 
  FrankNormalM & 6.60 & 6.70 & 5.80 & 5.00 & 3.60 & 3.90 & 5.80 & 6.10 & 3.50 & 4.40 \\ 
  ClaytonExponentialM & 6.10 & 5.00 & 6.00 & 6.00 & 3.60 & 5.00 & 5.30 & 4.60 & 3.70 & 4.80 \\ 
  ClaytonLinearM & 4.10 & 5.40 & 5.40 & 5.40 & 5.30 & 6.00 & 5.10 & 5.70 & 6.30 & 6.10 \\ 
  ClaytonNormalM & 5.30 & 4.30 & 4.80 & 4.40 & 5.10 & 4.70 & 4.90 & 5.60 & 5.90 & 5.50 \\ 
  GalambosExponentialM & 3.30 & 2.80 & 5.20 & 5.00 & 2.80 & 4.10 & 4.30 & 4.90 & 5.20 & 6.10 \\ 
  GalambosLinearM & 5.10 & 5.30 & 7.00 & 6.30 & 4.90 & 3.80 & 6.10 & 6.00 & 5.50 & 5.70 \\ 
  GalambosNormalM & 4.60 & 6.60 & 5.50 & 4.40 & 4.70 & 5.80 &  & 6.40 & 6.00 & 4.00 \\ 
  UniformFrankD5 & 4.30 & 4.70 & 5.40 & 5.10 & 4.00 & 5.00 & 4.70 & 4.70 & 6.70 & 5.00 \\ 
  FrankClaytonD5 & 5.10 & 6.30 & 5.60 & 6.10 & 6.00 & 4.20 & 6.60 & 6.30 & 4.30 & 5.20 \\ 
  FrankJoeD5 & 4.90 & 4.80 & 2.90 & 3.30 & 5.00 & 5.50 & 4.40 & 4.20 & 6.80 & 5.60 \\ 
  FrankNormalM5 & 3.80 & 4.20 & 5.50 & 4.90 & 4.60 & 4.40 & 4.20 & 4.20 & 4.20 & 6.00 \\ 
  FrankLinearM5 & 4.90 & 4.40 & 5.80 & 6.90 & 4.70 & 5.70 & 6.30 & 6.40 & 5.90 & 5.00 \\ 
  FrankExponentialM5 & 3.40 & 4.30 & 5.70 & 4.40 & 5.20 & 6.60 & 4.20 & 4.70 & 5.40 & 4.60 \\ 
  ClaytonExponentialM5 & 2.20 & 3.20 & 3.60 & 3.90 & 5.00 & 5.10 & 3.80 & 3.50 & 3.30 & 4.80 \\ 
  ClaytonLinearM5 & 4.70 & 4.40 & 3.40 & 3.50 & 5.20 & 5.90 & 5.50 & 5.60 & 6.30 & 4.90 \\ 
  ClaytonNormalM5 & 5.60 & 5.80 & 5.50 & 5.90 & 3.20 & 5.70 & 6.60 & 6.40 & 5.00 & 4.80 \\ 
  UniformExponentialM5 & 4.20 & 4.80 & 4.30 & 4.70 & 3.30 & 5.60 & 6.30 & 6.00 & 5.90 & 5.40 \\ 
   \hline
\end{tabular}
\end{table}

% latex table generated in R 4.5.1 by xtable 1.8-4 package
% Tue Jul 22 10:01:20 2025
\begin{table}[ht]
\centering
\begin{tabular}{rrrrrrrrr}
  \hline
 & NN0 & CF1 & CF2 & CF3 & CF4 & Ball & ES & EP \\ 
  \hline
NormalD2 & 5.60 & 5.50 & 4.00 & 5.50 & 4.60 & 5.90 & 4.90 & 4.70 \\ 
  tD2 & 6.30 & 4.20 & 4.10 & 4.20 & 5.40 & 5.30 & 4.10 & 5.80 \\ 
  UniformMixtureD2 & 6.50 & 5.10 & 4.60 & 5.10 & 4.60 & 4.80 & 4.60 & 3.90 \\ 
  FrankD2 & 7.00 & 5.30 & 5.50 & 5.30 & 6.00 & 4.90 & 4.20 & 3.90 \\ 
  ClaytonD2 & 5.70 & 5.40 & 4.70 & 5.40 & 5.60 & 6.00 & 4.30 & 3.50 \\ 
  GumbelD2 & 5.10 & 4.30 & 4.00 & 4.30 & 3.90 & 5.60 & 5.10 & 4.30 \\ 
  GalambosD2 & 4.80 & 5.30 & 5.10 & 5.30 & 5.30 & 5.20 & 4.30 & 5.10 \\ 
  HuslerReissD2 & 5.30 & 5.80 & 4.00 & 5.80 & 5.00 & 5.80 & 5.60 & 3.70 \\ 
  ClaytonGumbelD2 & 6.00 & 5.60 & 4.60 & 5.60 & 5.40 & 5.50 & 3.90 & 3.90 \\ 
  UniformFrankD2 & 5.50 & 5.30 & 4.60 & 5.30 & 4.40 & 4.90 & 3.30 & 2.70 \\ 
  ParetoSimplexD2 & 5.20 & 5.40 & 5.90 & 5.40 & 5.80 & 4.30 & 3.90 & 5.10 \\ 
  KhoudrajiClaytonD2 & 6.00 & 4.80 & 4.40 & 4.80 & 4.60 & 5.90 & 4.70 & 5.10 \\ 
  NormalUniformD2 & 5.50 & 5.20 & 5.20 & 5.20 & 5.60 & 4.40 & 4.80 & 4.40 \\ 
  JoeD2 & 6.40 & 4.60 & 5.00 & 4.60 & 4.10 & 4.60 & 3.60 & 4.30 \\ 
  DalitzCleoD2 & 4.70 & 4.70 & 4.60 & 4.70 & 5.20 & 5.00 & 5.10 & 5.20 \\ 
  DalitzPDGD2 & 5.40 & 3.80 & 4.60 & 3.80 & 4.60 & 6.60 & 5.20 & 4.30 \\ 
  DalitzBabarD2 & 6.50 & 5.30 & 4.70 & 5.30 & 4.30 & 5.70 & 6.00 & 4.30 \\ 
  NormalShiftM & 6.40 & 5.60 & 6.50 & 5.60 & 6.00 & 5.10 & 5.20 & 3.90 \\ 
  NormalStretchM & 6.60 & 5.20 & 4.30 & 5.20 & 4.60 & 3.60 & 3.60 & 4.30 \\ 
  UniformRotateM & 5.80 & 4.80 & 6.10 & 4.80 & 5.20 & 4.20 & 5.30 & 4.30 \\ 
  UniformBetaM & 7.00 & 5.80 & 5.20 & 5.80 & 5.40 & 6.10 & 3.70 & 3.20 \\ 
  TruncExponentialM & 5.70 & 5.10 & 5.40 & 5.10 & 6.00 & 5.20 & 4.30 & 5.20 \\ 
  DalitzCleoM & 4.20 & 5.00 & 5.20 & 5.00 & 5.10 & 6.50 & 5.80 & 6.30 \\ 
  DalitzPDGM & 6.30 & 4.80 & 6.20 & 4.80 & 5.20 & 4.80 & 5.20 & 4.80 \\ 
  DalitzBabarM & 4.30 & 3.90 & 4.80 & 3.90 & 4.20 & 5.30 & 4.10 & 3.00 \\ 
  NormalD5 & 6.10 & 5.00 & 5.10 & 5.00 & 4.30 & 3.90 &  &  \\ 
  tD5 & 6.00 & 5.00 & 3.60 & 5.00 & 4.60 & 4.60 &  &  \\ 
  FrankD5 & 6.20 & 5.70 & 4.60 & 5.70 & 4.60 & 6.60 &  &  \\ 
  ClaytonD5 & 5.90 & 3.90 & 5.60 & 3.90 & 4.40 & 6.10 &  &  \\ 
  GumbelD5 & 6.20 & 5.10 & 5.00 & 5.10 & 5.20 & 4.90 &  &  \\ 
  JoeD5 & 8.90 & 5.10 & 4.30 & 5.10 & 5.00 & 4.90 &  &  \\ 
  FrankExponentialM & 6.50 & 5.30 & 4.70 & 5.30 & 5.20 & 5.10 & 3.80 & 4.80 \\ 
  FrankLinearM & 6.30 & 5.80 & 5.20 & 5.80 & 5.50 & 5.20 & 4.80 & 4.10 \\ 
  FrankNormalM & 6.10 & 4.40 & 5.30 & 4.40 & 4.90 & 3.70 & 5.60 & 3.50 \\ 
  ClaytonExponentialM & 5.50 & 4.80 & 4.10 & 4.80 & 3.70 & 5.30 & 4.30 & 5.40 \\ 
  ClaytonLinearM & 6.30 & 6.10 & 5.30 & 6.10 & 6.10 & 5.50 & 3.70 & 4.80 \\ 
  ClaytonNormalM & 5.90 & 5.50 & 5.00 & 5.50 & 5.20 & 4.70 & 4.40 & 3.50 \\ 
  GalambosExponentialM & 5.90 & 6.10 & 4.70 & 6.10 & 5.00 & 5.80 & 3.90 & 4.70 \\ 
  GalambosLinearM & 5.60 & 5.70 & 4.70 & 5.70 & 4.60 & 5.70 & 5.40 & 5.10 \\ 
  GalambosNormalM & 6.40 & 4.00 & 4.10 & 4.00 & 5.00 & 4.90 & 4.90 & 4.90 \\ 
  UniformFrankD5 & 7.30 & 5.00 & 4.30 & 5.00 & 5.70 & 4.60 &  &  \\ 
  FrankClaytonD5 & 6.40 & 5.20 & 5.00 & 5.20 & 4.90 & 4.00 &  &  \\ 
  FrankJoeD5 & 6.60 & 5.60 & 5.40 & 5.60 & 5.20 & 5.00 &  &  \\ 
  FrankNormalM5 & 5.90 & 6.00 & 5.00 & 6.00 & 5.60 & 4.90 &  &  \\ 
  FrankLinearM5 & 5.60 & 5.00 & 6.20 & 5.00 & 5.70 & 5.10 &  &  \\ 
  FrankExponentialM5 & 6.10 & 4.60 & 5.80 & 4.60 & 5.90 & 4.70 &  &  \\ 
  ClaytonExponentialM5 & 6.10 & 4.80 & 5.40 & 4.80 & 6.30 & 4.20 &  &  \\ 
  ClaytonLinearM5 & 7.00 & 4.90 & 5.80 & 4.90 & 5.50 & 4.90 &  &  \\ 
  ClaytonNormalM5 & 7.50 & 4.80 & 4.20 & 4.80 & 5.20 & 4.00 &  &  \\ 
  UniformExponentialM5 & 6.30 & 5.40 & 4.80 & 5.40 & 5.00 & 5.10 &  &  \\ 
   \hline
\end{tabular}
\end{table}

\subsubsection{Continuous data - power estimates}

% latex table generated in R 4.5.1 by xtable 1.8-4 package
% Tue Jul 22 10:01:20 2025
\begin{table}[ht]
\centering
\begin{tabular}{rrrrrrrrrrr}
  \hline
 & KS & K & CvM & AD & NN1 & NN5 & AZ & BF & BG & FR \\ 
  \hline
NormalD2 & 42.00 & 43.00 & 55.00 & 64.00 & 48.00 & 66.00 & 88.00 & 83.00 & 5.00 & 60.00 \\ 
  tD2 & 40.00 & 39.00 & 54.00 & 64.00 & 36.00 & 50.00 & 82.00 & 78.00 & 7.00 & 49.00 \\ 
  UniformMixtureD2 & 24.00 & 25.00 & 34.00 & 42.00 & 48.00 & 68.00 & 91.00 & 65.00 & 6.00 & 63.00 \\ 
  FrankD2 & 29.00 & 26.00 & 41.00 & 46.00 & 22.00 & 29.00 & 84.00 & 72.00 & 4.00 & 32.00 \\ 
  ClaytonD2 & 88.00 & 97.00 & 93.00 & 98.00 & 99.00 & 100.00 & 100.00 & 100.00 & 7.00 & 100.00 \\ 
  GumbelD2 & 34.00 & 29.00 & 46.00 & 49.00 & 30.00 & 43.00 & 87.00 & 74.00 & 5.00 & 39.00 \\ 
  GalambosD2 & 38.00 & 35.00 & 46.00 & 54.00 & 31.00 & 45.00 & 89.00 & 76.00 & 6.00 & 39.00 \\ 
  HuslerReissD2 & 73.00 & 85.00 & 82.00 & 86.00 & 89.00 & 98.00 & 100.00 & 100.00 & 7.00 & 96.00 \\ 
  ClaytonGumbelD2 & 31.00 & 30.00 & 32.00 & 38.00 & 65.00 & 78.00 & 92.00 & 72.00 & 45.00 & 76.00 \\ 
  UniformFrankD2 & 34.00 & 36.00 & 39.00 & 40.00 & 44.00 & 64.00 & 82.00 & 59.00 & 32.00 & 60.00 \\ 
  ParetoSimplexD2 & 97.00 & 60.00 & 98.00 & 98.00 & 49.00 & 80.00 & 100.00 & 100.00 & 50.00 & 72.00 \\ 
  KhoudrajiClaytonD2 & 43.00 & 55.00 & 47.00 & 56.00 & 74.00 & 92.00 & 99.00 & 97.00 & 13.00 & 86.00 \\ 
  NormalUniformD2 & 36.00 & 34.00 & 44.00 & 52.00 & 37.00 & 49.00 & 84.00 & 75.00 & 4.00 & 42.00 \\ 
  JoeD2 & 43.00 & 38.00 & 49.00 & 48.00 & 40.00 & 59.00 & 94.00 & 84.00 & 5.00 & 56.00 \\ 
  DalitzCleoD2 & 66.00 & 22.00 & 83.00 & 82.00 & 69.00 & 86.00 & 85.00 & 64.00 & 15.00 & 82.00 \\ 
  DalitzPDGD2 & 58.00 & 18.00 & 75.00 & 72.00 & 78.00 & 91.00 & 88.00 & 60.00 & 12.00 & 85.00 \\ 
  DalitzBabarD2 & 74.00 & 34.00 & 88.00 & 84.00 & 76.00 & 92.00 & 90.00 & 72.00 & 24.00 & 84.00 \\ 
  NormalShiftM & 95.00 & 87.00 & 94.00 & 93.00 & 24.00 & 36.00 & 98.00 & 99.00 & 24.00 & 29.00 \\ 
  NormalStretchM & 24.00 & 55.00 & 20.00 & 40.00 & 18.00 & 23.00 & 63.00 & 57.00 & 90.00 & 24.00 \\ 
  UniformRotateM & 41.00 & 78.00 & 34.00 & 54.00 & 62.00 & 82.00 & 94.00 & 94.00 & 26.00 & 72.00 \\ 
  UniformBetaM & 66.00 & 91.00 & 57.00 & 78.00 & 31.00 & 45.00 & 91.00 & 86.00 & 100.00 & 38.00 \\ 
  TruncExponentialM & 82.00 & 47.00 & 89.00 & 87.00 & 13.00 & 15.00 & 85.00 & 88.00 & 46.00 & 16.00 \\ 
  DalitzCleoM & 49.00 & 40.00 & 66.00 & 93.00 & 58.00 & 79.00 & 94.00 & 70.00 & 76.00 & 70.00 \\ 
  DalitzPDGM & 37.00 & 37.00 & 44.00 & 79.00 & 47.00 & 70.00 & 89.00 & 70.00 & 94.00 & 56.00 \\ 
  DalitzBabarM & 74.00 & 46.00 & 70.00 & 93.00 & 55.00 & 77.00 & 95.00 & 84.00 & 96.00 & 66.00 \\ 
  NormalD5 & 74.00 & 32.00 & 96.00 & 96.00 & 38.00 & 78.00 & 78.00 & 66.00 & 6.00 & 46.00 \\ 
  tD5 & 72.00 & 33.00 & 95.00 & 96.00 & 42.00 & 74.00 & 60.00 & 40.00 & 6.00 & 46.00 \\ 
  FrankD5 & 59.00 & 24.00 & 85.00 & 91.00 & 48.00 & 88.00 & 91.00 & 78.00 & 6.00 & 61.00 \\ 
  ClaytonD5 & 50.00 & 17.00 & 92.00 & 98.00 & 56.00 & 91.00 & 94.00 & 82.00 & 6.00 & 61.00 \\ 
  GumbelD5 & 70.00 & 40.00 & 90.00 & 89.00 & 56.00 & 89.00 & 95.00 & 83.00 & 7.00 & 57.00 \\ 
  JoeD5 & 79.00 & 45.00 & 89.00 & 82.00 & 46.00 & 84.00 & 93.00 & 80.00 & 4.00 & 55.00 \\ 
  FrankExponentialM & 91.00 & 89.00 & 80.00 & 81.00 & 18.00 & 32.00 & 92.00 & 94.00 & 27.00 & 28.00 \\ 
  FrankLinearM & 62.00 & 60.00 & 49.00 & 49.00 & 8.00 & 13.00 & 74.00 & 78.00 & 6.00 & 12.00 \\ 
  FrankNormalM & 36.00 & 83.00 & 22.00 & 46.00 & 22.00 & 37.00 & 47.00 & 29.00 & 86.00 & 34.00 \\ 
  ClaytonExponentialM & 93.00 & 97.00 & 74.00 & 72.00 & 60.00 & 81.00 & 98.00 & 98.00 & 18.00 & 76.00 \\ 
  ClaytonLinearM & 58.00 & 71.00 & 37.00 & 39.00 & 45.00 & 62.00 & 92.00 & 88.00 & 7.00 & 58.00 \\ 
  ClaytonNormalM & 37.00 & 89.00 & 18.00 & 38.00 & 78.00 & 92.00 & 87.00 & 56.00 & 78.00 & 89.00 \\ 
  GalambosExponentialM & 62.00 & 70.00 & 41.00 & 40.00 & 62.00 & 77.00 & 87.00 & 80.00 & 10.00 & 73.00 \\ 
  GalambosLinearM & 20.00 & 62.00 & 14.00 & 19.00 & 62.00 & 81.00 & 64.00 & 39.00 & 10.00 & 74.00 \\ 
  GalambosNormalM & 28.00 & 72.00 & 19.00 & 32.00 & 69.00 & 86.00 & 72.00 & 37.00 & 56.00 & 76.00 \\ 
  UniformFrankD5 & 33.00 & 14.00 & 67.00 & 82.00 & 19.00 & 58.00 & 52.00 & 37.00 & 5.00 & 32.00 \\ 
  FrankClaytonD5 & 21.00 & 17.00 & 34.00 & 54.00 & 50.00 & 86.00 & 39.00 & 14.00 & 27.00 & 61.00 \\ 
  FrankJoeD5 & 46.00 & 78.00 & 39.00 & 44.00 & 57.00 & 83.00 & 37.00 & 19.00 & 12.00 & 61.00 \\ 
  FrankNormalM5 & 41.00 & 58.00 & 23.00 & 23.00 & 20.00 & 43.00 & 62.00 & 42.00 & 99.00 & 25.00 \\ 
  FrankLinearM5 & 46.00 & 29.00 & 60.00 & 62.00 & 13.00 & 31.00 & 82.00 & 86.00 & 6.00 & 19.00 \\ 
  FrankExponentialM5 & 83.00 & 59.00 & 82.00 & 84.00 & 18.00 & 42.00 & 86.00 & 87.00 & 68.00 & 26.00 \\ 
  ClaytonExponentialM5 & 78.00 & 62.00 & 82.00 & 81.00 & 18.00 & 39.00 & 91.00 & 92.00 & 62.00 & 21.00 \\ 
  ClaytonLinearM5 & 49.00 & 31.00 & 59.00 & 63.00 & 15.00 & 30.00 & 77.00 & 78.00 & 5.00 & 21.00 \\ 
  ClaytonNormalM5 & 28.00 & 54.00 & 16.00 & 20.00 & 22.00 & 46.00 & 55.00 & 35.00 & 99.00 & 28.00 \\ 
  UniformExponentialM5 & 60.00 & 36.00 & 65.00 & 50.00 & 11.00 & 21.00 & 77.00 & 77.00 & 76.00 & 13.00 \\ 
   \hline
\end{tabular}
\end{table}

% latex table generated in R 4.5.1 by xtable 1.8-4 package
% Tue Jul 22 10:01:20 2025
\begin{table}[ht]
\centering
\begin{tabular}{rrrrrrrrr}
  \hline
 & NN0 & CF1 & CF2 & CF3 & CF4 & Ball & ES & EP \\ 
  \hline
NormalD2 & 40.00 & 60.00 & 38.00 & 60.00 & 50.00 & 30.00 & 97.00 & 98.00 \\ 
  tD2 & 31.00 & 49.00 & 31.00 & 49.00 & 40.00 & 20.00 & 86.00 & 93.00 \\ 
  UniformMixtureD2 & 35.00 & 63.00 & 44.00 & 63.00 & 54.00 & 31.00 & 89.00 & 85.00 \\ 
  FrankD2 & 22.00 & 32.00 & 16.00 & 32.00 & 24.00 & 43.00 & 78.00 & 78.00 \\ 
  ClaytonD2 & 97.00 & 100.00 & 99.00 & 100.00 & 100.00 & 100.00 & 100.00 & 100.00 \\ 
  GumbelD2 & 25.00 & 39.00 & 22.00 & 39.00 & 31.00 & 39.00 & 82.00 & 82.00 \\ 
  GalambosD2 & 24.00 & 39.00 & 22.00 & 39.00 & 31.00 & 41.00 & 84.00 & 85.00 \\ 
  HuslerReissD2 & 80.00 & 96.00 & 89.00 & 96.00 & 94.00 & 98.00 & 100.00 & 100.00 \\ 
  ClaytonGumbelD2 & 67.00 & 76.00 & 69.00 & 76.00 & 72.00 & 24.00 & 91.00 & 89.00 \\ 
  UniformFrankD2 & 36.00 & 60.00 & 43.00 & 60.00 & 52.00 & 24.00 & 97.00 & 96.00 \\ 
  ParetoSimplexD2 & 53.00 & 72.00 & 52.00 & 72.00 & 63.00 & 100.00 & 96.00 & 93.00 \\ 
  KhoudrajiClaytonD2 & 60.00 & 86.00 & 69.00 & 86.00 & 78.00 & 59.00 & 99.00 & 99.00 \\ 
  NormalUniformD2 & 25.00 & 42.00 & 23.00 & 42.00 & 31.00 & 43.00 & 84.00 & 84.00 \\ 
  JoeD2 & 35.00 & 56.00 & 37.00 & 56.00 & 47.00 & 53.00 & 93.00 & 93.00 \\ 
  DalitzCleoD2 & 48.00 & 82.00 & 72.00 & 82.00 & 77.00 & 19.00 & 86.00 & 63.00 \\ 
  DalitzPDGD2 & 49.00 & 85.00 & 77.00 & 85.00 & 82.00 & 27.00 & 86.00 & 63.00 \\ 
  DalitzBabarD2 & 51.00 & 84.00 & 76.00 & 84.00 & 79.00 & 47.00 & 79.00 & 55.00 \\ 
  NormalShiftM & 23.00 & 29.00 & 16.00 & 29.00 & 21.00 & 97.00 & 84.00 & 76.00 \\ 
  NormalStretchM & 19.00 & 24.00 & 14.00 & 24.00 & 20.00 & 76.00 & 59.00 & 38.00 \\ 
  UniformRotateM & 58.00 & 72.00 & 51.00 & 72.00 & 63.00 & 92.00 & 77.00 & 24.00 \\ 
  UniformBetaM & 26.00 & 38.00 & 28.00 & 38.00 & 32.00 & 99.00 & 75.00 & 68.00 \\ 
  TruncExponentialM & 11.00 & 16.00 & 7.00 & 16.00 & 10.00 & 88.00 & 45.00 & 39.00 \\ 
  DalitzCleoM & 47.00 & 70.00 & 49.00 & 70.00 & 61.00 & 54.00 & 54.00 & 89.00 \\ 
  DalitzPDGM & 38.00 & 56.00 & 37.00 & 56.00 & 48.00 & 72.00 & 79.00 & 68.00 \\ 
  DalitzBabarM & 42.00 & 66.00 & 41.00 & 66.00 & 54.00 & 86.00 & 66.00 & 92.00 \\ 
  NormalD5 & 18.00 & 46.00 & 41.00 & 46.00 & 45.00 & 21.00 &  &  \\ 
  tD5 & 17.00 & 46.00 & 34.00 & 46.00 & 41.00 & 12.00 &  &  \\ 
  FrankD5 & 31.00 & 61.00 & 49.00 & 61.00 & 55.00 & 50.00 &  &  \\ 
  ClaytonD5 & 25.00 & 61.00 & 59.00 & 61.00 & 62.00 & 61.00 &  &  \\ 
  GumbelD5 & 24.00 & 57.00 & 54.00 & 57.00 & 57.00 & 49.00 &  &  \\ 
  JoeD5 & 21.00 & 55.00 & 51.00 & 55.00 & 55.00 & 40.00 &  &  \\ 
  FrankExponentialM & 20.00 & 28.00 & 16.00 & 28.00 & 22.00 & 75.00 & 78.00 & 72.00 \\ 
  FrankLinearM & 12.00 & 12.00 & 7.00 & 12.00 & 9.00 & 62.00 & 34.00 & 35.00 \\ 
  FrankNormalM & 30.00 & 34.00 & 24.00 & 34.00 & 30.00 & 30.00 & 80.00 & 58.00 \\ 
  ClaytonExponentialM & 58.00 & 76.00 & 56.00 & 76.00 & 67.00 & 68.00 & 94.00 & 99.00 \\ 
  ClaytonLinearM & 40.00 & 58.00 & 37.00 & 58.00 & 47.00 & 45.00 & 89.00 & 86.00 \\ 
  ClaytonNormalM & 75.00 & 89.00 & 77.00 & 89.00 & 83.00 & 21.00 & 94.00 & 82.00 \\ 
  GalambosExponentialM & 55.00 & 73.00 & 52.00 & 73.00 & 64.00 & 26.00 & 91.00 & 88.00 \\ 
  GalambosLinearM & 53.00 & 74.00 & 51.00 & 74.00 & 64.00 & 23.00 & 33.00 & 44.00 \\ 
  GalambosNormalM & 64.00 & 76.00 & 59.00 & 76.00 & 68.00 & 13.00 & 90.00 & 74.00 \\ 
  UniformFrankD5 & 16.00 & 32.00 & 22.00 & 32.00 & 27.00 & 16.00 &  &  \\ 
  FrankClaytonD5 & 11.00 & 61.00 & 82.00 & 61.00 & 80.00 & 10.00 &  &  \\ 
  FrankJoeD5 & 14.00 & 61.00 & 76.00 & 61.00 & 75.00 & 16.00 &  &  \\ 
  FrankNormalM5 & 58.00 & 25.00 & 65.00 & 25.00 & 64.00 & 80.00 &  &  \\ 
  FrankLinearM5 & 15.00 & 19.00 & 10.00 & 19.00 & 15.00 & 69.00 &  &  \\ 
  FrankExponentialM5 & 12.00 & 26.00 & 21.00 & 26.00 & 23.00 & 83.00 &  &  \\ 
  ClaytonExponentialM5 & 11.00 & 21.00 & 19.00 & 21.00 & 21.00 & 83.00 &  &  \\ 
  ClaytonLinearM5 & 15.00 & 21.00 & 12.00 & 21.00 & 15.00 & 71.00 &  &  \\ 
  ClaytonNormalM5 & 59.00 & 28.00 & 66.00 & 28.00 & 68.00 & 82.00 &  &  \\ 
  UniformExponentialM5 & 7.00 & 13.00 & 10.00 & 13.00 & 13.00 & 83.00 &  &  \\ 
   \hline
\end{tabular}
\end{table}

\subsubsection{Discrete data - type I error}

% latex table generated in R 4.5.1 by xtable 1.8-4 package
% Tue Jul 22 10:01:44 2025
\begin{table}[ht]
\centering
\begin{tabular}{rrrrrrrrr}
  \hline
 & KS & K & CvM & AD & NN & AZ & BF & Chisquare \\ 
  \hline
NormalD2 & 3.00 & 3.50 & 2.40 & 2.70 & 3.30 & 3.80 & 4.00 & 3.30 \\ 
  tD2 & 3.40 & 4.00 & 4.80 & 4.50 & 4.80 & 4.10 & 4.40 & 4.50 \\ 
  UniformMixtureD2 & 4.40 & 5.30 & 4.50 & 5.20 & 3.80 & 3.20 & 3.30 & 3.80 \\ 
  FrankD2 & 3.60 & 5.30 & 3.60 & 4.30 & 2.80 & 4.60 & 4.50 & 4.90 \\ 
  ClaytonD2 & 3.40 & 3.90 & 4.80 & 5.00 & 2.40 & 4.70 & 4.40 & 4.10 \\ 
  GumbelD2 & 3.70 & 3.80 & 4.10 & 4.10 & 3.30 & 5.50 & 5.70 & 5.10 \\ 
  GalambosD2 & 4.20 & 3.00 & 4.00 & 4.00 & 1.80 & 4.90 & 4.70 & 3.50 \\ 
  HuslerReissD2 & 3.30 & 5.00 & 5.10 & 4.60 & 2.50 & 4.30 & 4.60 & 5.90 \\ 
  ClaytonGumbelD2 & 2.90 & 2.90 & 4.00 & 4.10 & 2.90 & 4.40 & 4.30 & 4.30 \\ 
  UniformFrankD2 & 4.30 & 3.90 & 5.10 & 4.20 & 3.40 & 4.30 & 4.00 & 3.90 \\ 
  ParetoSimplexD2 & 3.60 & 4.50 & 4.30 & 4.50 & 3.60 & 3.60 & 5.20 & 4.10 \\ 
  KhoudrajiClaytonD2 & 4.10 & 3.90 & 3.90 & 3.90 & 3.50 & 4.30 & 5.20 & 4.10 \\ 
  NormalUniformD2 & 4.00 & 2.20 & 3.80 & 3.80 & 2.00 & 4.30 & 4.90 & 3.70 \\ 
  JoeD2 & 3.70 & 3.60 & 4.10 & 4.20 & 3.70 & 5.20 & 6.00 & 4.20 \\ 
  DalitzCleoD2 & 3.80 & 2.40 & 3.20 & 3.30 & 3.00 & 3.30 & 3.80 & 5.50 \\ 
  DalitzPDGD2 & 3.40 & 3.40 & 3.50 & 3.90 & 3.30 & 3.80 & 3.70 & 4.10 \\ 
  DalitzBabarD2 & 2.80 & 4.00 & 3.90 & 4.40 & 1.30 & 4.20 & 4.20 & 3.40 \\ 
  NormalShiftM & 2.10 & 3.60 & 3.90 & 4.10 & 3.00 & 3.90 & 4.40 & 4.40 \\ 
  NormalStretchM & 3.70 & 4.00 & 4.00 & 4.40 & 2.80 & 4.50 & 3.80 & 4.40 \\ 
  UniformRotateM & 5.00 & 3.80 & 4.00 & 4.60 & 2.00 & 4.50 & 3.80 & 5.00 \\ 
  UniformBetaM & 4.40 & 2.70 & 4.40 & 4.20 & 3.60 & 5.30 & 3.80 & 6.20 \\ 
  TruncExponentialM & 4.80 & 5.10 & 4.40 & 3.60 & 3.90 & 6.20 & 4.90 & 4.40 \\ 
  DalitzCleoM & 3.90 & 3.70 & 5.80 & 4.50 & 2.50 & 3.80 & 4.00 & 5.10 \\ 
  DalitzPDGM & 2.80 & 2.90 & 4.10 & 2.70 & 3.60 & 3.90 & 3.80 & 4.40 \\ 
  DalitzBabarM & 3.90 & 4.50 & 4.20 & 5.70 & 2.80 & 4.10 & 4.80 & 5.70 \\ 
  FrankExponentialM & 3.60 & 2.60 & 5.10 & 5.20 & 3.90 & 2.80 & 5.10 & 4.30 \\ 
  FrankLinearM & 2.70 & 2.90 & 3.10 & 3.10 & 2.60 & 4.00 & 3.90 & 4.90 \\ 
  FrankNormalM & 4.40 & 4.00 & 5.50 & 4.70 & 4.50 & 2.90 & 5.40 & 4.40 \\ 
  ClaytonExponentialM & 3.40 & 3.90 & 3.50 & 3.10 & 3.10 & 3.30 & 3.40 & 5.50 \\ 
  ClaytonLinearM & 2.70 & 3.10 & 3.90 & 3.70 & 2.60 & 3.40 & 3.50 & 5.10 \\ 
  ClaytonNormalM & 2.80 & 3.10 & 3.70 & 3.50 & 1.70 & 3.80 & 3.70 & 4.60 \\ 
  GalambosExponentialM & 2.80 & 3.30 & 3.30 & 3.40 & 3.80 & 4.20 & 3.90 & 3.00 \\ 
  GalambosLinearM & 3.60 & 3.10 & 3.60 & 4.20 & 4.30 & 5.20 & 4.10 & 5.10 \\ 
  GalambosNormalM & 2.60 & 3.60 & 3.40 & 3.10 & 3.20 & 3.70 & 5.10 & 4.30 \\ 
   \hline
\end{tabular}
\end{table}

\subsubsection{Discrete data - power estimates}

% latex table generated in R 4.5.1 by xtable 1.8-4 package
% Tue Jul 22 10:01:44 2025
\begin{table}[ht]
\centering
\begin{tabular}{rrrrrrrrr}
  \hline
 & KS & K & CvM & AD & NN & AZ & BF & Chisquare \\ 
  \hline
NormalD2 & 27.00 & 24.00 & 32.00 & 38.00 & 44.00 & 4.00 & 4.00 & 97.00 \\ 
  tD2 & 22.00 & 23.00 & 18.00 & 21.00 & 38.00 & 4.00 & 4.00 & 88.00 \\ 
  UniformMixtureD2 & 23.00 & 24.00 & 23.00 & 30.00 & 79.00 & 6.00 & 6.00 & 89.00 \\ 
  FrankD2 & 25.00 & 24.00 & 33.00 & 39.00 & 48.00 & 4.00 & 3.00 & 77.00 \\ 
  ClaytonD2 & 81.00 & 98.00 & 88.00 & 94.00 & 100.00 & 4.00 & 4.00 & 100.00 \\ 
  GumbelD2 & 24.00 & 20.00 & 33.00 & 36.00 & 38.00 & 5.00 & 5.00 & 83.00 \\ 
  GalambosD2 & 23.00 & 27.00 & 34.00 & 36.00 & 34.00 & 4.00 & 4.00 & 84.00 \\ 
  HuslerReissD2 & 67.00 & 80.00 & 75.00 & 80.00 & 94.00 & 5.00 & 4.00 & 100.00 \\ 
  ClaytonGumbelD2 & 22.00 & 24.00 & 28.00 & 33.00 & 84.00 & 6.00 & 5.00 & 92.00 \\ 
  UniformFrankD2 & 27.00 & 30.00 & 34.00 & 33.00 & 32.00 & 6.00 & 4.00 & 96.00 \\ 
  ParetoSimplexD2 & 95.00 & 52.00 & 97.00 & 98.00 & 71.00 & 98.00 & 98.00 & 95.00 \\ 
  KhoudrajiClaytonD2 & 32.00 & 42.00 & 36.00 & 44.00 & 86.00 & 4.00 & 4.00 & 99.00 \\ 
  NormalUniformD2 & 26.00 & 31.00 & 32.00 & 40.00 & 48.00 & 4.00 & 3.00 & 83.00 \\ 
  JoeD2 & 39.00 & 38.00 & 36.00 & 34.00 & 52.00 & 6.00 & 5.00 & 93.00 \\ 
  DalitzCleoD2 & 25.00 & 9.00 & 46.00 & 50.00 & 62.00 & 14.00 & 19.00 & 67.00 \\ 
  DalitzPDGD2 & 42.00 & 11.00 & 55.00 & 55.00 & 76.00 & 8.00 & 8.00 & 84.00 \\ 
  DalitzBabarD2 & 55.00 & 10.00 & 68.00 & 66.00 & 64.00 & 27.00 & 24.00 & 83.00 \\ 
  NormalShiftM & 95.00 & 84.00 & 90.00 & 93.00 & 55.00 & 37.00 & 37.00 & 82.00 \\ 
  NormalStretchM & 24.00 & 51.00 & 23.00 & 31.00 & 33.00 & 6.00 & 4.00 & 60.00 \\ 
  UniformRotateM & 54.00 & 82.00 & 52.00 & 61.00 & 37.00 & 17.00 & 57.00 & 76.00 \\ 
  UniformBetaM & 48.00 & 79.00 & 42.00 & 51.00 & 53.00 & 22.00 & 4.00 & 74.00 \\ 
  TruncExponentialM & 74.00 & 34.00 & 84.00 & 84.00 & 29.00 & 55.00 & 76.00 & 41.00 \\ 
  DalitzCleoM & 13.00 & 13.00 & 18.00 & 24.00 & 25.00 & 12.00 & 4.00 & 38.00 \\ 
  DalitzPDGM & 34.00 & 49.00 & 40.00 & 57.00 & 36.00 & 12.00 & 12.00 & 70.00 \\ 
  DalitzBabarM & 42.00 & 43.00 & 50.00 & 59.00 & 39.00 & 5.00 & 5.00 & 62.00 \\ 
  FrankExponentialM & 85.00 & 89.00 & 75.00 & 79.00 & 51.00 & 90.00 & 29.00 & 78.00 \\ 
  FrankLinearM & 56.00 & 57.00 & 39.00 & 37.00 & 18.00 & 3.00 & 16.00 & 36.00 \\ 
  FrankNormalM & 32.00 & 70.00 & 22.00 & 50.00 & 40.00 & 98.00 & 71.00 & 80.00 \\ 
  ClaytonExponentialM & 88.00 & 94.00 & 58.00 & 65.00 & 69.00 & 92.00 & 36.00 & 94.00 \\ 
  ClaytonLinearM & 58.00 & 69.00 & 31.00 & 31.00 & 65.00 & 5.00 & 25.00 & 90.00 \\ 
  ClaytonNormalM & 30.00 & 77.00 & 17.00 & 45.00 & 64.00 & 92.00 & 50.00 & 95.00 \\ 
  GalambosExponentialM & 48.00 & 62.00 & 30.00 & 33.00 & 63.00 & 67.00 & 22.00 & 90.00 \\ 
  GalambosLinearM & 23.00 & 34.00 & 13.00 & 13.00 & 24.00 & 3.00 & 9.00 & 33.00 \\ 
  GalambosNormalM & 20.00 & 59.00 & 7.00 & 31.00 & 57.00 & 75.00 & 25.00 & 90.00 \\ 
   \hline
\end{tabular}
\end{table}

\bibliographystyle{apalike}
\bibliography{references}

\end{document}